%% file: ms.tex
\pdfoutput=1
\documentclass[3p]{elsarticle}

\usepackage[utf8]{inputenc}
\usepackage{caption}
\usepackage{hyperref}
\usepackage{mathtools}
\usepackage{amsfonts}
\usepackage{float}
\usepackage{xparse}
\usepackage[nofillcomment,linesnumbered,vlined,boxed,commentsnumbered]{algorithm2e}
\usepackage{multirow}
\usepackage{xfrac}
\usepackage{listings}
\usepackage{afterpage}
\usepackage[table]{xcolor}
\usepackage{tikz}
\usepackage{pdfpages}

\input{commands}

\begin{document}
\input{front}

\input{introduction}
\input{problem_statement}
\input{system_architecture}
\input{algorithms}

\input{estimation}
\input{experiments}
\input{conclusion}
\input{acknowledgment}

\section*{References}
\bibliography{bibfile} 

\appendix
\input{appendix}

\end{document}

%% file: commands.tex
\newtheorem{thm}{Theorem}
\newtheorem{lem}{Lemma}
\newproof{pf}{Proof}

\DeclareMathOperator*{\argmin}{arg\,min}
\DeclarePairedDelimiter{\ceil}{\lceil}{\rceil}
\newcommand\abs[1]{\left|#1\right|}

\DeclareDocumentCommand \tasked { O{\task} m } {{#2}^{(#1)}}

\DeclareDocumentCommand \resourced { O{\resource} m } {{#2}^{(#1)}}

\DeclareDocumentCommand \taskedresourced { O{\task} O{\resource} m } {
  {#3}^{(#1, #2)}
}

\DeclareDocumentCommand \taskedsol { O{\task} O{\probsol} m } {
    {#3}_{\probsol}^{(#1)}
}

\DeclareDocumentCommand \taskify { m m } {
  \DeclareDocumentCommand #1 { s o } {
    \IfBooleanTF {##1} {
      \tasked{#2}
    }{
      \IfNoValueTF {##2} {
        #2
      }{
        \tasked[##2]{#2}
      }
    }
  }
}

\DeclareDocumentCommand \resourcify { m m } {
  \DeclareDocumentCommand #1 { s o } {
    \IfBooleanTF {##1} {
      \resourced{#2}
    }{
      \IfNoValueTF {##2} {
        #2
      }{
        \resourced[##2]{#2}
      }
    }
  }
}

\DeclareDocumentCommand \taskifysolufy { m m } {
  \DeclareDocumentCommand #1 { s o o } {
    \IfBooleanTF {##1} {
      \taskedsol{#2}
    }{
      \IfNoValueTF {##2} {
        \IfNoValueTF {##3} {
          #2
        }{
        }
      }{
        \IfNoValueTF {##3} {
          \taskedsol[##2]{#2}
        }{
          \taskedsol[##2][##3]{#2}
        }
      }
    }
  }
}

\newcommand{\dummyval}{\infty}

\newcommand{\naturalset}{{\mathbb{N}}}
\newcommand{\realset}{{\mathbb{R}}}
\newcommand{\prealset}{{\mathbb{R}_{>0}}}
\newcommand{\pnaturalset}{{\mathbb{N}_{>0}}}

\newcommand{\unifdist}{\mathcal{U}}

\newcommand{\vx}{{\mathbf{x}}}
\newcommand{\vy}{{\mathbf{y}}}

\newcommand{\np}{{\mathcal{NP}}}

\newcommand{\define}[1]{{\textit{#1}}}
\newcommand{\highlight}[1]{{\textbf{#1}}}

\newcommand{\task}{{i}}

\newcommand{\resource}{{j}}

\newcommand{\timeset}{T}

\taskify{\arrival}{a}

\taskify{\weight}{w}

\newcommand{\schprob}{\Pi}

\newcommand{\taskset}{N}
\resourcify{\ntasks}{n}
\newcommand{\npendtasks}{\ntasks[*]}

\newcommand{\resourceset}{M}
\newcommand{\nresources}{m}

\newcommand{\tasktypeset}{\Omega}
\taskify{\tasktype}{\omega}

\taskify{\reqresp}{rrt}

\taskify{\reqdue}{rdd}

\taskify{\critcurve}{Z}

\taskify{\critcurvet}{t}

\taskify{\critcurvev}{z}

\taskify{\critcurveq}{\q}

\taskify{\qualcurve}{Q}

\newcommand{\instanceset}{\Psi}
\taskify{\instance}{\psi}

\newcommand{\nptfn}{p}

\newcommand{\enptfn}{\widehat{\nptfn}}

\newcommand{\nwcttimefn}{WCT}

\newcommand{\fitparam}{y}
\newcommand{\fitparams}{\vy}
\taskify{\afitparams}{\widetilde{\fitparams}}
\taskify{\efitparams}{\widehat{\fitparams}}
\taskify{\afitparam}{\widetilde{\fitparam}}
\taskify{\efitparam}{\widehat{\fitparam}}

\taskify{\features}{\vx}

\newcommand{\q}{\phi}

\newcommand{\qset}{\Phi}

\newcommand{\minq}{\underline{\q}}

\newcommand{\qfn}{q}

\newcommand{\qfnsoltup}{\qfn_{\probsol}}

\taskifysolufy{\qfnsol}{\qfn}

\newcommand{\ralloc}{r}

\taskifysolufy{\rallocsol}{\ralloc}

\newcommand{\rallocsoltup}{\ralloc_{\probsol}}

\newcommand{\alloc}{st}

\taskifysolufy{\allocsol}{\alloc}

\newcommand{\allocsoltup}{\alloc_{\probsol}}

\newcommand{\ape}{ape}

\resourcify{\speed}{s}

\resourcify{\espeed}{\widehat{s}}

\resourcify{\speedfn}{s}

\resourcify{\espeedfn}{\widehat{s}}

\resourcify{\capab}{\tasktypeset}

\resourcify{\connectfn}{\delta}

\resourcify{\benchres}{bt}

\newcommand{\modelfn}{g}

\newcommand{\compfn}{ct}

\newcommand{\ecompfn}{\widehat{\compfn}}

\newcommand{\solqfn}{\varphi}

\newcommand{\probsol}{S}
\newcommand{\probsolset}{\mathcal{S}}

\newcommand{\avesolqfn}{\overline{\varphi}}

\newcommand{\avenlatfn}{\overline{L}}

\newcommand{\lat}{L}

\newcommand{\featurefn}{f}

\newcommand{\featdim}{D}

%% file: front.tex
\begin{frontmatter}
\includepdf[pages=1,fitpaper,noautoscale]{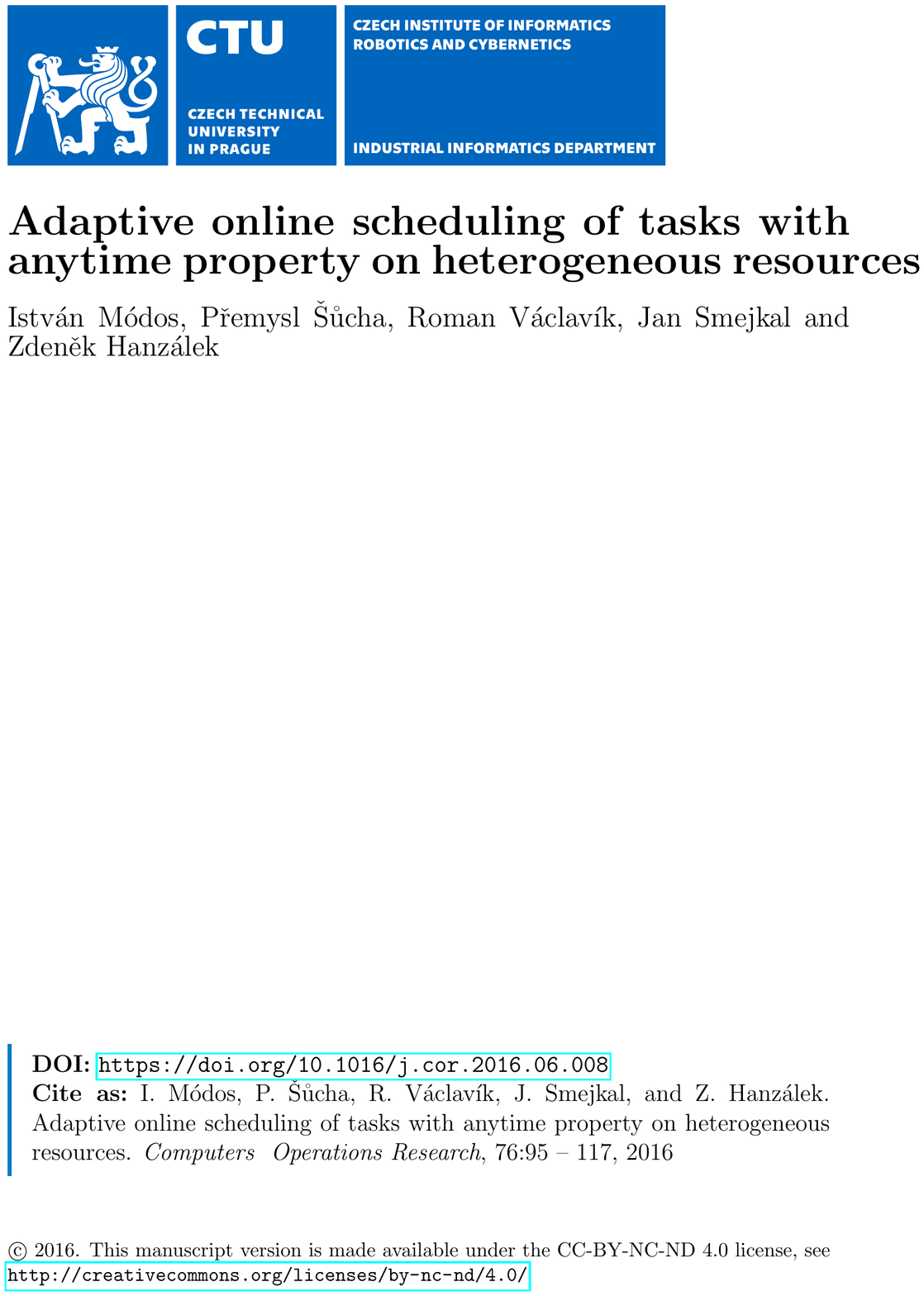}

\title{Adaptive online scheduling of tasks with anytime property on heterogeneous resources}

\author[cvut]{István Módos}
\ead{modosist@fel.cvut.cz}
\author[cvut]{Přemysl Šůcha}
\ead{suchap@fel.cvut.cz}
\author[cvut]{Roman Václavík}
\ead{vaclarom@fel.cvut.cz}
\author[merica]{Jan Smejkal}
\ead{smejkal@merica.cz}
\author[cvut]{Zdeněk Hanzálek}
\ead{hanzalek@fel.cvut.cz}

\address[cvut]{Department of Control Engineering, Faculty of Electrical Engineering, Czech Technical University, Karlovo náměstí 13, 121 35 Prague 2, Czech Republic}
\address[merica]{Merica, U Ládek 353/37, 251 01 Říčany – Strašín, Czech Republic}

\begin{abstract}
An acceptable response time of a server is an important aspect in many client-server applications;
this is evident in situations in which the server is overloaded by many computationally intensive requests.
In this work, we consider that the requests, or in this case \define{tasks}, generated by the clients are instances of optimization problems solved by anytime algorithms, i.e. the quality of the solution increases with the processing time of a task.
These tasks are submitted to the server which schedules them to the available computational resources where the tasks are processed.
To tackle the overload problem, we propose a scheduling algorithm which combines traditional scheduling approaches with a quality control heuristic which adjusts the requested quality of the solutions and thus changes the processing time of the tasks.
Two efficient quality control heuristics are introduced: the first heuristic sets a global quality for all tasks, whereas the second heuristic sets the quality for each task independently.
Moreover, in practice, the relationship between the processing time and the quality is not known \textit{a priori}.
Because it is crucial for scheduling algorithms to know at least the estimation of these relationships, we propose a general procedure for estimating these relationships using information obtained from the already executed tasks.
Finally, the performance of the proposed scheduling algorithm is demonstrated on a real-world problem from the domain of personnel rostering with very good results.
\end{abstract}

\begin{keyword}
Online scheduling, anytime algorithms, machine learning, adaptive systems
\end{keyword}

\end{frontmatter}

%% file: introduction.tex
\section{Introduction}
\label{sec:intro}
An important aspect of client-server applications is the response time of the server.
In a case of computationally intensive requests, e.g. optimization problems, the issue of the response time is even more pressing because the server can be easily overwhelmed even by a small number of requests.

Due to financial reasons, the computational capacity of a server is commonly scaled to handle a typical workload, i.e. the arrival rate and the computational complexity of the requests, so that the response time during this typical workload is kept at an acceptable level.
In a case of sudden increase in the requests, the server may become easily overloaded and the response time increases significantly resulting in user dissatisfaction.
One possibility of how to mitigate the increased response time during the overload is to buy more computational resources, but such solution is not financially suitable if the overload occurs a few times a day.
However, if the requests or some of the requests are instances of optimization problems, it is possible to maintain an acceptable response time by moderate degradation of the solution quality, i.e. to trade-off a small decrease in a solution quality for a significantly shorter response time.

\begin{figure}[H]
  \centering
  \begin{minipage}[t]{.5\textwidth}
    \centering
    \includegraphics[scale=0.45]{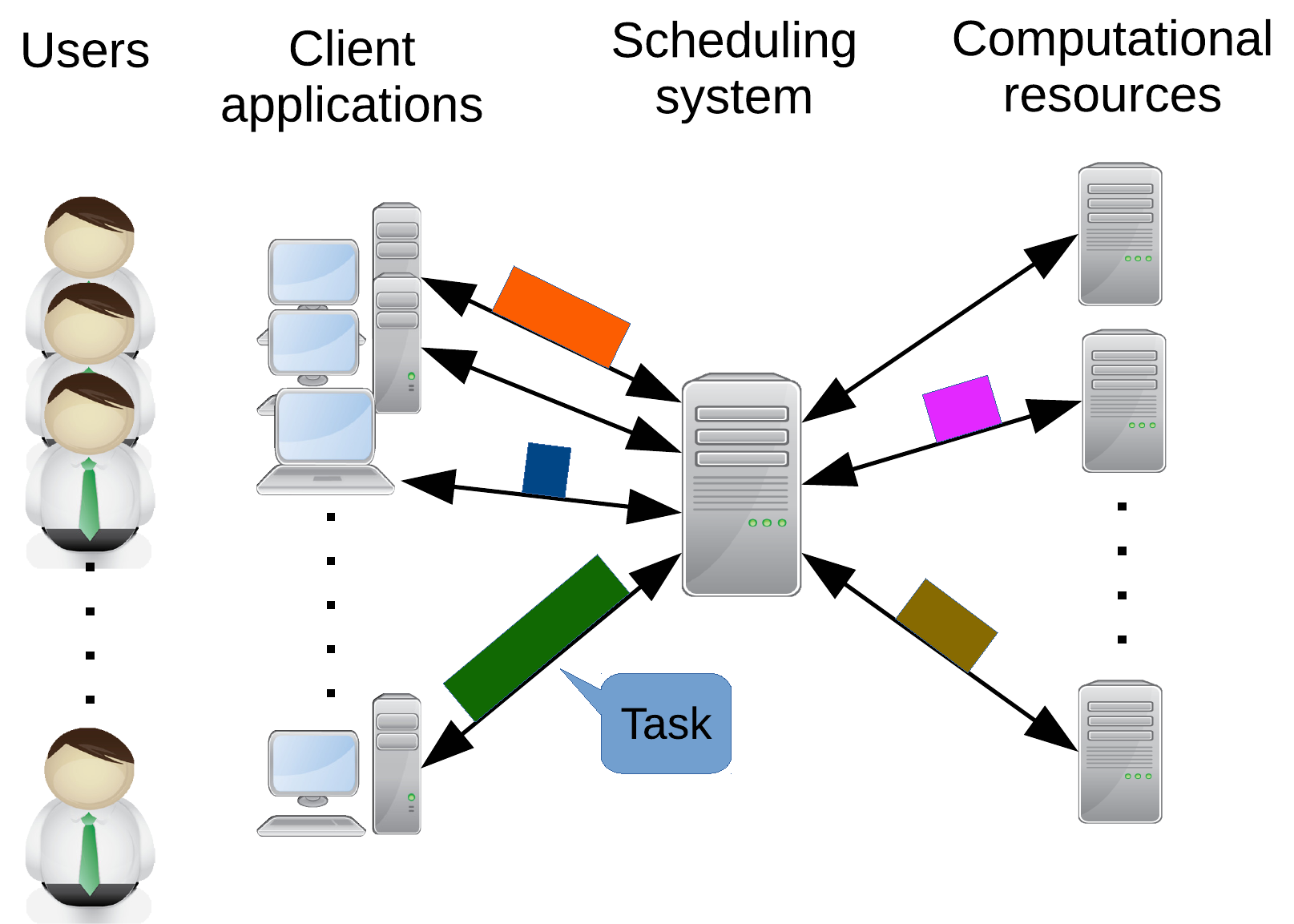}
    \captionof{figure}{Overview of the environment.}
    \label{fig:intro/arch}
  \end{minipage}%
  \begin{minipage}[t]{.5\textwidth}
    \centering
    \includegraphics[scale=0.35]{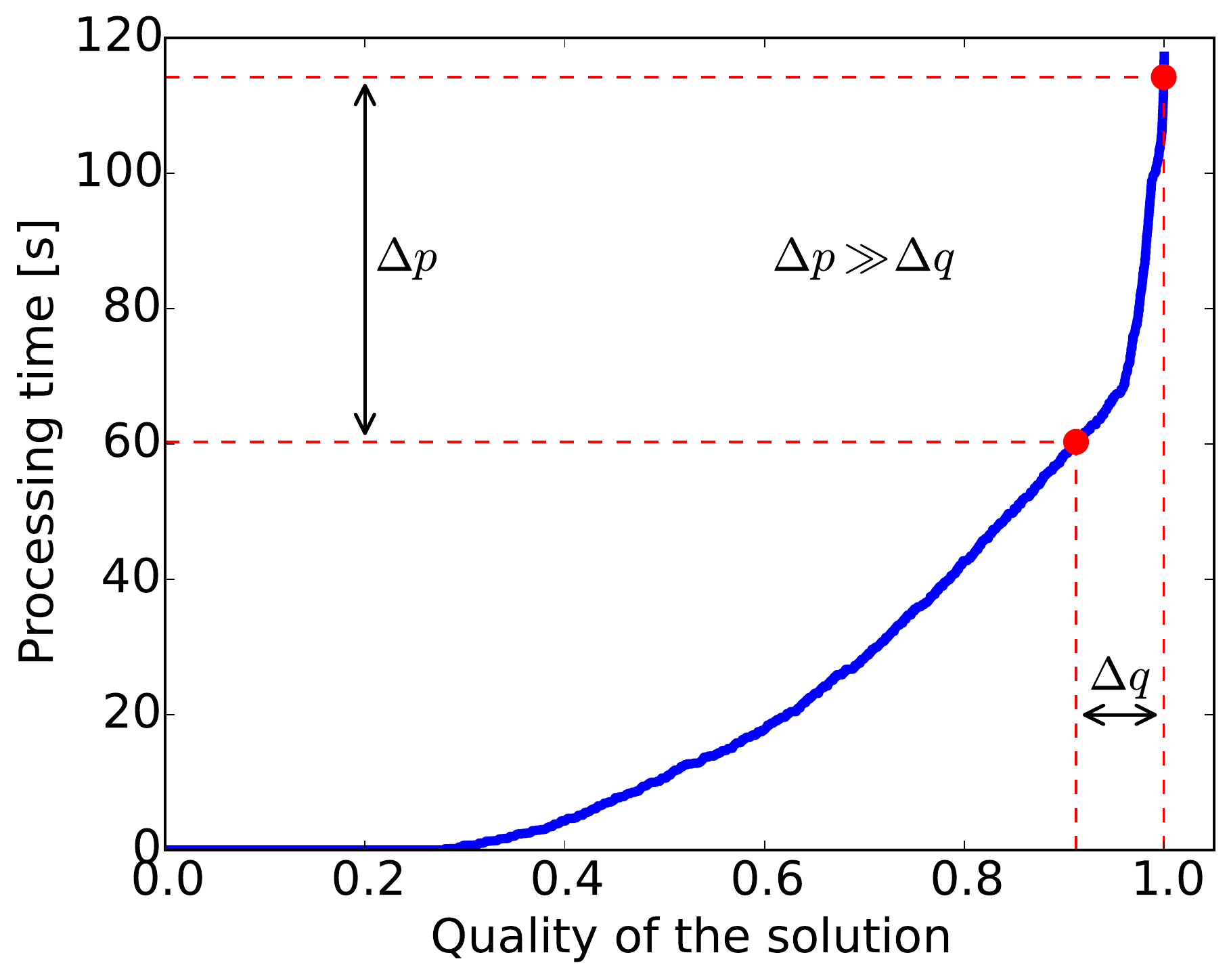}
    \captionof{figure}{A typical example of the processing time function of one task.}
    \label{fig:intro/response}
  \end{minipage}
\end{figure}

In this paper, we consider a scheduling problem illustrated in Figure \ref{fig:intro/arch}.
\define{Users} work with \define{client applications} which generate \define{tasks}.
The tasks are sent to a \define{scheduling system} which schedules the received tasks to \define{computational resources}.
The resources are \define{heterogeneous}, i.e. each resource may have a different processing power and, therefore, the processing time of the tasks may vary on each resource.
The task is processed on the assigned resource, and once it is finished, its result is sent back to the scheduling system which distributes the result to the respective client application.
The scheduling system receives the tasks progressively through time, i.e. we are dealing with an \define{online scheduling} problem \cite{hoogeveen1996,sgall1998a}.

The tasks are instances of some optimization problems and are solved by \define{anytime algorithms}.
The property of an anytime algorithm is that the processing of a task can be interrupted at any time and the algorithm returns a feasible solution if such solution exists.
The quality of the solution depends on the processing time of a task, i.e. a longer processing time may result in a better solution (the quality of a solution is defined using the objective function of the tasks' optimization problem, i.e. how close the solution is to the optimal/near optimal solution).
This behavior is typical for the majority of metaheuristics and hyperheuristics solving optimization problems.
The relationships between the processing time and the solution quality are defined by \define{processing time functions}.
A typical example of a processing time function of one task is illustrated in \hbox{Figure \ref{fig:intro/response}}.
In general, these functions have an increasing character: to get a better solution, an anytime algorithm must perform more operations or explore a larger part of the solution space. 
From Figure \ref{fig:intro/response}, it can also be seen that a slight deterioration of the solution quality can significantly shorten the processing time of the task and thus reduce the response time of the system.
From the user point of view, a good solution is better then excessive waiting time for the near-optimal/optimal solution.

In reality, the processing time functions are not known \textit{a priori} as is usually considered in the related literature.
The reason for this is that the anytime algorithms search the solution space and the algorithms are not generally aware where the good solutions are.
Without any knowledge of the processing time functions, the scheduling system cannot guarantee the response time because the scheduling system does not know how long the processing of the tasks will take.
However, using either statistical or machine learning methods, the processing time for the given quality can be estimated from the previous executions of similar tasks.

In this study, we focus on the situations in which the scheduling system is \define{overloaded}, i.e. the response time of the system increases significantly due to increased workload.
The idea of how to tackle the system overload is to control the requested quality of solutions, i.e. when the overload of the system is detected, the system trades off the quality of solutions so that the response time is kept close to the acceptable level.
On the other hand, the system requests the highest quality of the solutions if the overload is not detected. 

We want to emphasize that our solution does not substitute \define{clouds} \cite{foster2008}.
In fact, a cloud can be integrated into the scheduling system as a cluster of computational resources.
However, the proposed scheduling system replaces the default cloud load balancer because the scheduling policy (see Section \ref{sec:alg/sched}) makes more informed decisions than the default cloud load balancer and thus increases the responsiveness more efficiently. For example, Amazon Elastic Load Balancer uses simple Round-Robin algorithm for TCP listeners and Least-Number-Of-Requests for HTTP listeners  \cite{awsbalancer2015}. 

\subsection{Case study}
\label{sec:intro/case_study}
The validity of the proposed scheduling system is evaluated on an existing web-based personnel rostering application called Roslab\footnote{\url{http://www.merica.cz/products/roslab/benefits}}.
\define{Personnel rostering} \cite{ernst2004} is a combinatorial optimization problem where a set of shifts is assigned to employees so that \define{hard constraints} (typically given by the labor code) are satisfied and the penalty accumulated due to \define{soft constraints} violations (representing the quality of a solution) is minimized.
An assignment of shifts to employees is called a \define{roster}.

The main functionality provided by the Roslab client application is: (i) rostering, i.e. an automatic design of a roster, (ii) rerostering, i.e. a partial roster correction due to external events such as illness of an employee, and (iii) validation of the roster changes which were introduced by users.
Different user interactions with the client application generate different computational intensive tasks (e.g. the automatic roster design) and these tasks are solved by the anytime heuristic algorithms \cite{baumelt2014} on the server side.

At present, the available computational capacity of Roslab is scaled to handle the average workload.
However, the response time increases considerably when many users interact with the system at the same time, which causes a higher risk of not prolonging the service contract.
Because these critical situations occur only a few times a day, it is not financially suitable to buy new computational resources.
To increase the responsiveness, we can exploit the fact that the algorithms implemented in Roslab for solving the tasks have typically non-linear progress of the solution quality, i.e. the quality difference between the optimal and a high-quality solution is small while the processing time difference is significant (see Figure \ref{fig:intro/response}).
Therefore, it is possible to apply the proposed algorithms to control the trade-off between the solution quality and responsiveness.

\subsection{Related work}
\subsubsection{Scheduling and overload}
In distributed systems such as grids \cite{fangpeng2006a}, most practical scheduling problems are $\np$-hard \cite{xhafa2010a}.
The schedulers must also consider and balance different objective functions, e.g. response time, the fairness of resource sharing \cite{klusacek2014}, load balancing, etc.
Therefore, most schedulers use some heuristics which strive to find a reasonable solution to the given problem in an acceptable amount of time, i.e. the solution does not have to be optimal, but it should be found quickly.
Various heuristics, such as \define{Min-Min} \cite{ibarra1977}, \define{Sufferage} \cite{maheswaran1999}, \define{Local search} \cite{braun2001,xhafa2010a} or \define{Genetic algorithms} \cite{braun2001,xhafa2010a} have been presented.
A good scheduling algorithm can utilize all available resources and, therefore, handle many tasks in parallel.
However, in the case of overload when the capacity of resources is not enough, additional methods must be employed.
Buying a new computational resource is reasonable if the system is constantly overloaded.
However, if the overload is sparse, the purchase of a new resource is not justified because the resource would just consume energy without actually processing any tasks.

A common approach to handling the overload is to employ an \define{admission control} \cite{welsh2003a, gilly2012} which drops the tasks of some users when the overload is detected to guarantee the acceptable response time for the rest of the users.
However, such a solution is not acceptable in our case because all users must be served.

Another approach is to shorten the processing time of each task so that some performance objective is maximized.
The control of the processing times has different names in the literature: \textit{imprecise computation} \cite{liu1991a}, \textit{controllable processing time} \cite{shabtay2007, tseng2009}, \textit{partial jobs} \cite{chin2003}, \textit{increasing reward with increasing service} \cite{dey1996a}, and \textit{Quality of Service degradation} \cite{mittal1998a, he2011}.
The closest work to our problem is \cite{tseng2009}, where an algorithm for balancing the tardiness and total cost of the processing time compression on a single machine is given.
The relationship between the compression cost and the processing time of the tasks is linear.
Although the presented algorithm seems to have a good asymptotic complexity of $\mathcal{O}(n^2)$, where $n$ is the number of the tasks, we think that the actual complexity should be $\mathcal{O}(p_{max} \cdot n^2)$ where $p_{max}$ is the maximum processing time of the tasks.
The reason for this is that the presented algorithm in each iteration decreases the processing time of some task by 1 and, in the worst case, the algorithm will iterate until the processing time of the tasks cannot be further compressed.
The authors probably considered the maximum processing time as a constant - in this case, the former complexity would be correct.
Therefore, the scheduling algorithm is not suitable for the online environment where the range of the processing times is large.
In \cite{he2011}, the relationship between the solution quality and the processing time is assumed to be increasing and concave.
This assumption is reasonable for problems solved by anytime algorithms.
However, the presented algorithm has a high complexity of $\mathcal{O}(n^4)$ and the authors assume that all tasks share the same quality profile function.
Other works allow preemption \cite{liu1991a,dey1996a,chin2003} which is not suitable because if the tasks consume a lot of memory, then context switching could incur a big overhead.
Another assumption used in the cited works is that the processing time of the tasks is known or that the due dates are strict \cite{liu1991a}.
In our environment, it is not critical to meet due dates for all tasks but for the majority of tasks, i.e. we deal with \define{soft real-time} scheduling. 

Our first idea for the quality control was to use \define{control theory} \cite{hellerstein2004a}.
An example of an application of the control theory in computing systems is \cite{chenyang1999a} where a combination of the admission control and Quality of Service degradation using a \define{Proportional-Integral-Derivative} controller is described.
However, the control theory assumes that the state of the controlled system is in the neighborhood of the operating point.
If the controlled system is non-linear and the state is far away from the operating point, the control could produce undesirable behavior such as oscillations.
As an example, consider a sudden increase in the arrival rate of tasks to the server which changes the effect of the actuators on the measured output considerably.
A possible approach to solving this problem is to adapt the parameters of the control law dynamically when a change in the system state is detected \cite{karlsson2005a}.
The disadvantage of the parameters adaptation approach is that it cannot adapt to rapid changes in a workload because it needs some time to acquire a sufficient number of past measurements before the parameters are adapted.
For the previously mentioned reason, we designed quality control algorithms which do not rely on the parameters describing the operating point of the system (see Section \ref{sec:alg/quality}).

\subsubsection{Estimation of the processing time functions}
Because the processing time functions of the tasks (see Figure \ref{fig:intro/response}) are not known \textit{a priori}, they have to be estimated.
Some simple methods were proposed in the literature such as using the average of the last $n$ values of the processing time.
However, these methods cannot be used if the processing times of the tasks are significantly different.
A more successful approach is to use statistical methods or methods from machine learning \cite{page2008, hutter2014}.
In comparison with single resource environment, estimation of the processing time of the tasks in heterogeneous systems is even more complicated, i.e. the same task may have different processing times on the resources with a different processing power.
Therefore, the same estimation cannot be directly used for different resources.
In \cite{iverson1999}, this problem is solved by \define{benchmarking} the resources.
When a new resource connects to the scheduling system, series of benchmarking tests are run on this resource.
The result of these tests is a vector of numbers where each number denotes how well the resource performed on the corresponding test.
The authors assume that the processing time of one particular task is the same on resources with similar benchmark results and, therefore, they use observations of the processing time on one resource to estimate the processing time on a different resource.
In \cite{iverson1999}, the benchmark result is appended to a feature vector of a task and the $k$-nearest neighbor method is used to find similar observations from which an estimation of the processing time is computed.
An alternative approach to appending a benchmark result to a feature vector is to scale the processing time of a task by the benchmark result of the resource on which the task is processed \cite{page2008}.
However, none of these works deal with scheduling of tasks solved by anytime algorithms.

\subsection{Contribution and outline}
From the summarized related work it can be seen that no work fully addresses the problem of online scheduling of computationally intensive tasks which are solved by anytime algorithms and for which the processing time functions are not known beforehand.
The existing literature either assumes exact knowledge or specific shape of the processing time functions which is not realistic in practice.
Moreover, some works allow preemption \cite{liu1991a,dey1996a,chin2003}, optimize a different objective function of the scheduling problem \cite{tseng2009} or the presented algorithms have a high complexity \cite{he2011}.
Therefore, we introduce a new modular scheduling system which can guarantee an acceptable response time even in the case of overload.
Moreover, since the processing time functions of the tasks are not known \define{a priori}, we propose an estimator which can provide the estimation of the whole processing time function using knowledge acquired from the execution of similar tasks.
The proposed scheduling system is evaluated on a real client-server application from the domain of personnel rostering.

Our approach to tackling the overload problem combines a traditional scheduling policy with a quality control algorithm.
We propose two novel and efficient quality control algorithms: (i) \define{bisection control}, which sets one global quality for all tasks and (ii) \define{independent control}, which controls the quality of the solutions for each task independently.
The experiments in Section \ref{sec:exp} show that the proposed scheduling policy and quality control algorithms: (i) can decrease the response time significantly in situations in which the system would be overloaded if the quality control was disabled, (ii) outperform a simple control approach which always stops the task that was running for the longest time, and (iii) are robust to small errors in the estimation of the processing time functions.

The rest of the paper is organized as follows: In Section \ref{sec:prob} we formulate our scheduling problem formally.
The overview of the scheduling system is presented in Section \ref{sec:arch}.
Section \ref{sec:alg} contains the description of the presented scheduling policy and quality control algorithms.
Section \ref{sec:estim} explains how the estimation of the processing time functions works.
In Section \ref{sec:exp}, the proposed system is experimentally tested on real world instances.
Finally, the last section concludes the paper.

%% file: problem_statement.tex
\section{Problem statement}
\label{sec:prob}
In the whole text, it is assumed that the time is discrete and that one time unit equals one millisecond.
$\timeset = \naturalset$ denotes a set of time instants $t$.
The \define{quality of solutions} (or just quality) is denoted as $\q \in \qset$, where $\qset = \left[0, 1\right]$.
Interval $\qset$ can be understood as normalized values of the tasks' objective functions, i.e. 1 represents the best possible solution while 0 represents the worst possible solution.

The scheduling problem is defined by a tuple $\schprob = \left(\taskset, \resourceset,\instanceset, \featurefn, \nptfn, \nwcttimefn\right)$, where $\taskset = \{1, \dots, \ntasks\}$ is a set of \define{tasks}, $\resourceset = \{1, \dots, \nresources\}$ is a set of heterogeneous \define{resources} on which the tasks are processed, $\instanceset$ is a set of all possible \define{instances}, $\featurefn: \instanceset \rightarrow \realset^{\featdim}$ represents a \define{feature function}, which maps each \define{instance} $\instance \in \instanceset$ to a \define{feature vector} with dimensionality of $\featdim \in \pnaturalset$, $\nptfn: \instanceset \times \qset \rightarrow \pnaturalset$ represents a \define{normalized processing time function}, and $\nwcttimefn: \instanceset \rightarrow \pnaturalset$ is a \define{normalized worst case processing time function} (the meaning of the normalization will be explained later in this section).
The instances in this context are instances of the optimization problem solved by the computational resources, e.g. \textit{Personnel rostering} in the considered case study.
For arbitrary instance $\instance \in \instanceset$, it is assumed that $\nptfn(\instance, \q)$ is continuous and increasing function, i.e.
$\forall \q_1, \q_2 \in \qset:\,\, \q_1 < \q_2 \implies \nptfn(\instance, \q_1) < \nptfn(\instance, \q_2)$.
The definition allows the normalized processing time function to be non-linear in a general case.
A \define{maximum normalized processing time} of instance $\instance$ is a normalized processing time for quality 1.
A \define{normalized worst case processing time} $\nwcttimefn(\instance)$ denotes an upper bound of the processing time after which the anytime algorithm, solving the given task, is stopped.

In the whole text, notation $\task \in \taskset$ and $\resource \in \resourceset$ is used to denote the tasks and resources, respectively.
To highlight that an object is indexed by a task or a resource, we will use superscripts $(\task)$ or $(\resource)$, respectively.

Each task $\task \in \taskset$ is represented by a tuple $\left(\arrival*, \reqresp*, \instance* \right)$, where $\arrival* \in \timeset$ is an \define{arrival time} of the task, $\reqresp* \in \pnaturalset$ is a \define{requested response time} of the task, and $\instance* \in \instanceset$ is a \define{task instance}.
Roughly, a task is an instance which arrived in the scheduling system.
It is assumed that the tasks cannot be preempted.
Time $\reqdue* = \arrival* + \reqresp*$ represents a \define{requested due date} of task $\task$, i.e. a soft deadline.
The \define{response time} is defined as the duration between the arrival of the task to the scheduling system and its completion.
The requested response time represents the preferred maximum response time set by the client applications.

A parameter \define{speed} $\speed* \in \prealset$ is defined for each resource $\resource$.
The speed of resource $\resource$ can be understood as a constant processing power, i.e. it is the amount of work done by the algorithm running on resource $\resource$ per time unit.
For example, consider an algorithm consisting of one loop which multiplies two numbers in each iteration.
A resource with the speed of 1 can make only one iteration per time unit whereas a resource with the speed of 10 can make ten iterations per time unit.
By a \textit{normalized processing time} we mean a processing time abstracted from the heterogeneity of the resources, i.e. it represents the processing time on a resource with the speed of 1 and, therefore, it also represents the amount of work to be done to get a solution of the given quality on such a resource.

A \define{quality function} $\qfn: \timeset \rightarrow \qset$ assigns a value of the requested quality of solutions to each time $t \in \timeset$.
The quality function is not defined by the problem but is part of the decision made by the scheduling system.

If task $\task$ is started at time $t$ on resource $\resource$ for some quality function $\qfn$, its \define{completion time} is defined as
\begin{equation}
  \compfn(\task, \resource, t, \qfn) =
  \min
  \Big\{
    t_{min} :
    t_{min} \in \timeset,\, t_{min} \ge t,\,
    (t_{min} - t)\speed* \ge
    \nptfn(\instance*, \qfn(t_{min}))
  \Big\}\,.
\end{equation}
The definition can be understood as the shortest time when the amount of work undertaken by an algorithm is greater than or equal to the currently requested amount of work (the requested amount of work equals to the normalized processing time).
Note that this definition allows the quality function $\qfn$ to vary over time.
If this were not the case, then the completion time could be easily defined as $\ceil*{\frac{\nptfn(\instance*, \q)}{\speed*}} + t$, where $\q$ is a constant quality value.

The \define{lateness} of task $\task$ if started at time $t$ on resource $\resource$ for some quality function $\qfn$ is defined as
\begin{equation}
  \lat(\task, \resource, t, \qfn) = \compfn(\task, \resource, t, \qfn) -
  \reqdue*\,.
\end{equation}

The \define{solution quality} of task $\task$ if started at time $t$ on resource $\resource$ for some quality function $\qfn$ is defined as
\begin{equation}
  \label{eq:solqfn}
  \solqfn(\task, \resource, t, q) =
  \max
  \Big\{
    \q :
    \q \in \qset,\,
    \left(\compfn(\task, \resource, t, q) - t\right) \speed* \ge
    \nptfn(\instance*, \q)
  \Big\}\,.
\end{equation}
Again, this definition allows the quality to change over time.
The reason for the inequality is to handle the case when the requested amount of work for the maximum quality is less than the actual amount of work done.

A \define{solution} to problem $\schprob$ is a tuple $\probsol = \left(\rallocsoltup, \allocsoltup, \qfnsoltup\right)$, where $ \rallocsoltup = \left(\rallocsol[1], \rallocsol[2], \dots, \rallocsol[\ntasks]\right)$ is a vector in which $\rallocsol* \in \resourceset \cup \{\dummyval\}$ maps task $\task$ to some resource or $\dummyval$, $\allocsoltup = \left(\allocsol[1], \allocsol[2], \dots, \allocsol[\ntasks]\right)$ is a vector in which $\allocsol* \in \timeset \cup \{\dummyval\}$ maps task $\task$ to the start time or $\dummyval$, and $\qfnsoltup = \left(\qfnsol[1], \qfnsol[2], \dots, \qfnsol[\ntasks]\right)$ is a vector of quality functions for each task.
Value $\dummyval$ represents uninitialized value, i.e. $\rallocsol* = \dummyval$ denotes that task $\task$ is not assigned to any resource and $\allocsol*=\dummyval$ denotes that task $\task$ does not have a starting time in solution $\probsol$.
Solution $\probsol = (\rallocsoltup, \allocsoltup, \qfnsoltup)$ is \define{feasible} if the following conditions are satisfied for each task $\task \in \taskset$:
\begin{align}
  \label{eq:fsol_ralloc_def}
  \rallocsol* \not=\dummyval\,, \\
  \label{eq:fsol_alloc_def}
  \allocsol* \not=\dummyval\,, \\
  \label{eq:fsol_arr}
  \arrival* \le \allocsol* \,, \\
  \label{eq:fsol_one}
  \forall \task, \task' \in \taskset:
  \task \not= \task' \wedge \rallocsol* = \rallocsol[\task']
  \implies
    \compfn(\task, \rallocsol*, \allocsol*, \qfnsol*) \le \allocsol[\task']
    \vee
    \compfn(\task', \rallocsol[\task'], \allocsol[\task'], \qfnsol[\task']) \le \allocsol*
\end{align}
The constraints require that: a task is assigned to some resource \eqref{eq:fsol_ralloc_def}, \eqref{eq:fsol_alloc_def}; a task cannot start before its arrival time \eqref{eq:fsol_arr}; and processing of two tasks on the same resource cannot overlap \eqref{eq:fsol_one}.
The \define{set of all feasible solutions} is denoted as $\probsolset$.

For the feasible solutions the following functions are defined: \define{average solution quality} \eqref{eq:avesolq} and \define{average normalized lateness} \eqref{eq:nalat}
\begin{align}
  \label{eq:avesolq}
  \avesolqfn(\probsol) &= \frac{1}{\ntasks}\sum_{\task \in \taskset}
      \solqfn(\task,\rallocsol*, \allocsol*, \qfnsol*)\,, \\
  \label{eq:nalat}
  \avenlatfn(\probsol) &=\frac{1}{\ntasks}\sum_{\task \in \taskset}
  \frac{\lat(\task, \rallocsol*, \allocsol*, \qfnsol*)}
     {\reqresp*}
\end{align}

The goal in the scheduling problem is to find such a feasible solution which minimizes the average normalized lateness while the average solution quality is maximized, i.e.
\begin{equation}
  \begin{array}{rcl}
    \min &\quad &\left(\avenlatfn(\probsol), -\avesolqfn(\probsol)\right) \\
    \text{s.t.} &\quad &\probsol \in \probsolset \\
  \end{array}
\end{equation}
Because the described problem is a multi-objective scheduling problem \cite{hoogeveen2005}, the outcome is a set of solutions called \define{Pareto front}.
For each solution in the Pareto front holds that it is not dominated by any other solution in the Pareto front.
Since the scheduling system has to react in an automatic manner, it needs to select some solution from the Pareto front so that the desired behavior of the system is achieved.
Obviously, an increase in the average solution quality leads to increase in the average normalized lateness and vice-versa.
Solutions which only optimize the average solution quality or the average normalized lateness are clearly unacceptable since the other objective is ignored.
Therefore, solutions which balance the solution quality and the normalized lateness are sought.

One possible way of achieving the balance is to minimize the weighted sum of the objectives.
However, due to non-linearity of the processing time functions, such solutions may lead to decreased average solution quality even if the system is not overloaded.
Moreover, it is not obvious how to set the weights since the scale of the objectives is different.

We argue that if the tasks are computed within the requested response time, the client applications are sufficiently responsive from the users point of view.
Therefore, we propose to maximize the average quality such that, on average, the tasks are processed very close to their requested due date.
This requirement can be expressed by pushing the average normalized lateness as close to 0 as possible.
This aggregation is correct even if the requested response time is significantly larger than the time needed to find the near optimal solutions since we perform ``tail-cutting'' on the processing time functions (see Section \ref{sec:estim} for more details).

\subsection{Extension to online scheduling}
\label{sec:prob/online}

The problem described above considers the complete information about tasks and resources.
However, in our setting, some quantities are unknown to the scheduler until time $t$ or until some conditions are satisfied (this relates to the online scheduling paradigm where tasks arrive over time \cite{sgall1998a}): 
\begin{itemize}
  \item The values of function $\nptfn$ for instance $\instance*$ of task $\task$ can be observed only after task $\task$ has finished.

  \item The values $\left(\arrival*, \reqresp*, \instance* \right)$ of each task $i$ are unknown until its arrival time $\arrival*$.

  \item The completion time and the solution quality of task $\task$ are known only after task $\task$ has finished.
\end{itemize}

On the other hand, the scheduler has full knowledge of the following information: (i) normalized worst case processing time function $\nwcttimefn$, (ii) feature functions $\featurefn$, (iii) dimensionality $\featdim$ of the feature vectors, and (iv) set of resources $\resourceset$ and speed of each resource (the speed of the resources can be acquired through benchmarking \cite{iverson1999}).

%% file: system_architecture.tex
\section{Scheduling system overview}
\label{sec:arch}
\begin{figure}[t]
  \centering
  \includegraphics[scale=0.4]{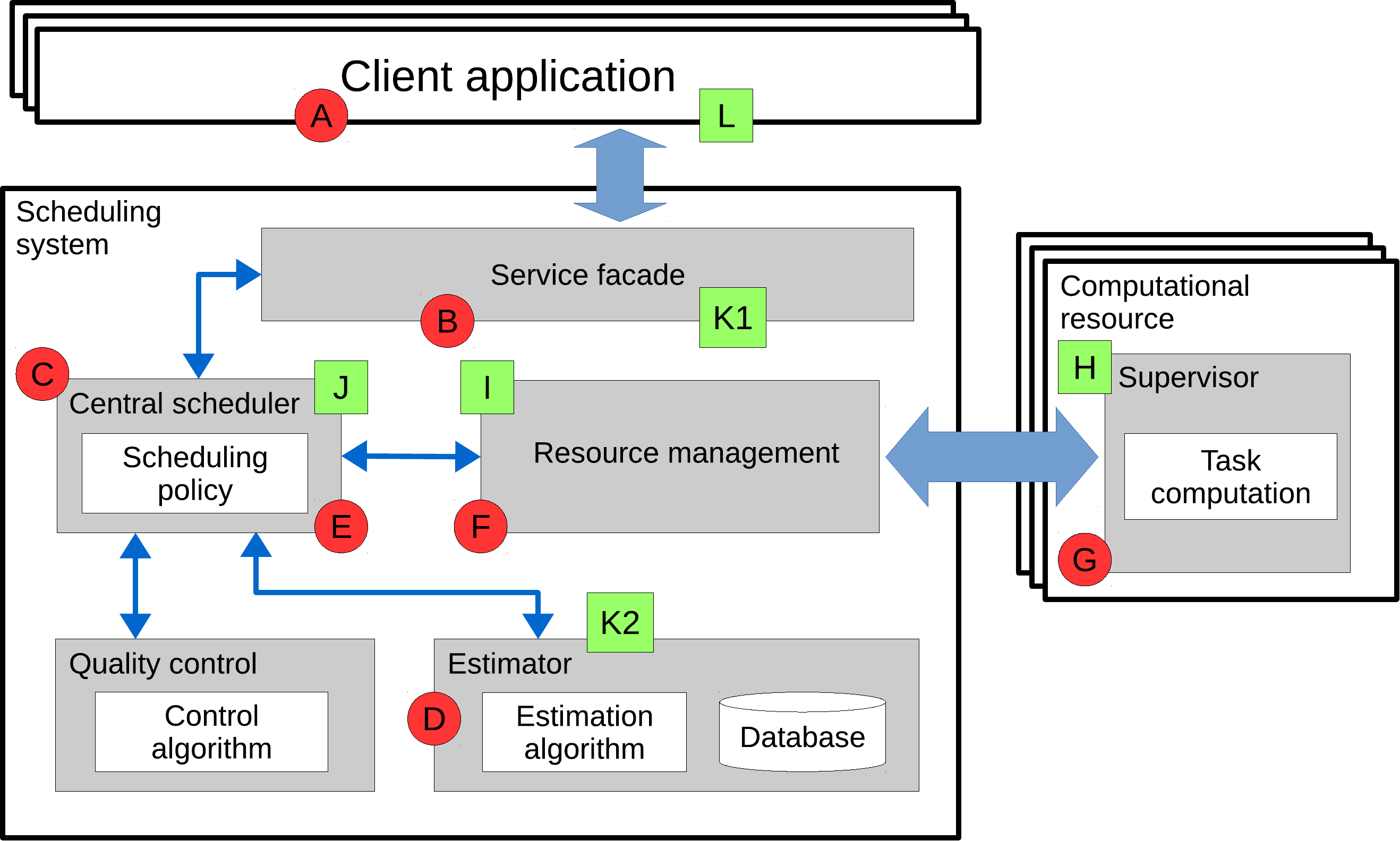}
  \caption{The proposed architecture.}
  \label{fig:arch/arch}
\end{figure}

With respect to the problem statement described in Section \ref{sec:prob}, we propose a modular architecture of the scheduling system.
Each module is implemented as an independent thread.
Therefore, the introduced architecture can better utilize the available computational capacity of the server on which the scheduling system is deployed.
The proposed architecture is depicted in Figure \ref{fig:arch/arch}.
It consists of three blocks:
\begin{enumerate}
  \item \define{Client applications}, which generate tasks and send them to the \define{scheduling system}.

  \item \define{Scheduling system}, which is responsible for assigning the received tasks from the \define{client applications} to the available \define{computational resources}.
    The \define{scheduling system} can be further divided into the following \define{modules}:
    \begin{itemize}
      \item \highlight{Service facade}, which acts as a communication interface between the scheduling system and the client applications. The interface allows the client applications to send tasks, abort tasks and to receive the solution of a previously sent task.

      \item \highlight{Central scheduler}, which stores the received tasks and assigns these tasks to the available resources using a \define{scheduling policy}.
        The data structure containing the information about the assignments of the tasks is called a \define{schedule}.
        The central scheduler also converts the requested quality of the solutions (given by the \highlight{quality control} module) to the actual processing time of each task and sends them to the computational resources through the \highlight{resource management}.

      \item \highlight{Resource management}, which monitors the computational resources and abstracts the communication between the central scheduler and the computational resources, e.g. communication over a socket with remote resources.

      \item \highlight{Estimator}, which estimates the normalized processing time functions, i.e. it provides $\enptfn: \instanceset \times \qset \rightarrow \pnaturalset$.
        These estimations are found by the \define{estimation algorithm} which uses the \define{regression model}.

      \item \highlight{Quality control}, which analyzes the current schedule created by the central scheduler and sets the requested quality of the solutions, i.e. it provides $\qfnsol*(t)$.
        The qualities are found by a \define{control algorithm}.
    \end{itemize}

  \item \define{Computational resources}, which process the assigned tasks and send their solutions back to the scheduling system.
    The computational resources can be local threads of the server, remote high-performance resources, etc.
    On each computational resource, a \define{supervisor} is executed which communicates with the scheduling system and is responsible for monitoring the computation of a task.
    We remind the reader that each resource can process only one task at a time.
\end{enumerate}

\subsection{Processing of a generated task}
To explain how the tasks flow through the system, Figure \ref{fig:arch/arch} is used.
The letters in red circles represent the order of the modules for an unprocessed task, whereas the letters in green rectangles represent the order of modules for a processed task.

When the \highlight{client application} (A) generates task $\task$, the \highlight{client application} sends that task to the \highlight{central scheduler} through the \highlight{service facade} (B).
The \highlight{central scheduler} (C) inserts that task to an array of \define{tasks to be estimated}, i.e. tasks for which the normalized processing time function needs to be estimated.
Then, the \highlight{central scheduler} sends that task to the \highlight{estimator} (D).
If a regression model is already trained, then the regression model is used to estimate the normalized processing time function for task $\task$.
If the regression model is not trained, the normalized processing time function is computed from the normalized worst case processing time.
The estimated normalized processing time function is provided to the \highlight{central scheduler} (E) which then moves task $\task$ from the array of tasks to be estimated to an array of \define{pending tasks}, i.e. tasks which were received by the scheduling system, are not completed yet and for which the normalized processing time function is already estimated.
The \highlight{central scheduler} creates a schedule of the pending tasks on the available resources using the scheduling policy.
When some resource $\resource$ becomes free, i.e. it is not processing any tasks, the scheduling system takes the first task from the schedule of the respective resource $\resource$ and assigns that task to that resource.
The task assignment is performed through the \highlight{resource management} (F) module which communicates with the computational resources.
When the task is received by the \highlight{supervisor} (G) of the resource, the \highlight{supervisor} spawns a new thread which executes the received task.
The processing time of task $\task$ is computed from the requested quality and the speed of resource $\resource$.
Since the processing time is computed from the requested quality which may change over time, resource $\resource$ is constantly notified about these changes.
The anytime algorithms on the computational resources run approximately for the duration of these processing times.

When the task has finished, the \highlight{supervisor} (H) sends the solution back to the \highlight{central scheduler} through the \highlight{resource management} (I).
The \highlight{central scheduler} (J) deletes task $\task$ from the array of pending tasks and sends the solution to the respective \highlight{client application} (L) through the \highlight{service facade} (K1).
If the progress of the solution quality over time was measured by the task algorithm, it is sent to the \highlight{estimator} (K2) which uses it to refine the regression model (see Section \ref{sec:estim/collect}).

%% file: algorithms.tex
\section{Scheduling policy and quality control algorithms}
\label{sec:alg}
If we consider the offline version of our problem without the quality control, then the closest scheduling problem can be represented in Graham's notation \cite{brucker2007} as $Qm|r_\task|\sum w_\task L_\task$.
This problem considers heterogeneous resources with quantifiable speed and that the tasks arrive in the system at \define{release time} $r_\task$ which is equal to the arrival time $\arrival*$ from our problem statement in Section \ref{sec:prob}.
The aim is to find a solution which minimizes the weighted lateness where the weight is defined as $w_\task = {}^{1}/_{\reqresp*}$.
The optimal solution to problem $Qm|r_\task|\sum w_\task L_\task$ is equal to the optimal solution to problem $Qm|r_\task|\sum w_\task C_\task$, the only difference is in the constant in the objective functions.
Unfortunately, even the simpler problem $1|r_\task|\sum w_\task C_\task$ is $\np$-hard \cite{brucker2007}, therefore, the problem $Qm|r_\task|\sum w_\task L_\task$ is also $\np$-hard.

For the online version (again, without the quality control), the situation is even more complicated \cite{sgall1998a}; this is due to the lack of knowledge of the tasks that will arrive later from the client application.
In such a case, the online algorithm might assign a longer task on an available resource, even though it would be more beneficial to wait if a shorter task arrives very soon.
Without prediction of the arrival of future tasks, an offline scheduling rule can be employed to create a schedule of all pending tasks.

Our approach to handling the online scheduling with the quality control is to divide this problem into two steps performed by the central scheduler and the quality control module:
\begin{enumerate}
  \item Central scheduler: For the maximum quality of the solutions, find a schedule of all pending tasks which minimizes the average normalized lateness.
    The schedule is found by the scheduling policy and is recreated whenever some \define{event} occurs (see Section \ref{sec:alg/sched} for a list of events).

  \item Quality control module: For a fixed schedule, find the requested quality of the solution for each task such that the average normalized lateness is close to 0.
    The quality control procedure is started whenever a new schedule is created by the central scheduler.
\end{enumerate}

To describe the algorithms, some additional notation needs to be introduced:
\begin{itemize}
  \item In algorithms, an \define{array} data structure is used.
    An array is an ordered sequence of some objects.
    Bracket notation is used for accessing the elements of an array, e.g. if $a = (1,3,5)$ is an array, then $a[3]=5$.
    The length of an array is computed using a $len$ function, e.g. the length of the array $a$ is $len(a) = 3$.
    To append an element to the end of an array, bracket notation is used with $end+1$ as the index, e.g. after calling $a[end +1] \leftarrow 7$, the content of $a$ will be $(1,3,5,7)$.

  \item An \define{estimated completion time} of task $\task$ started at time $t$ on resource $\resource$ with fixed quality $\q$ is defined as
    \begin{equation}
      \label{eq:ecompfn}
      \ecompfn(\task, \resource, t, \q) = t +
      \ceil*{\frac{\enptfn(\instance*, \q)}{\speedfn*}}\,.
    \end{equation}

  \item The number of pending tasks is denoted as $\npendtasks$ and the number of pending tasks assigned to resource $\resource$ is denoted as $\ntasks[\resource]$.
\end{itemize}

\subsection{Central scheduler}
\label{sec:alg/sched}
The policy used for scheduling is described in Algorithm \ref{alg:alg/edd_mct}.
It is a combination of an \define{Earliest Due Date} (EDD) selection policy and a \define{Minimal Completion Time} (MCT) assignment policy.
Although there are other policies, such as \define{First In First Out} and \define{Shortest Task First}, the EDD policy is chosen because it considers the requested due date and it does not suffer from starvation of longer tasks.

The global variables of the central scheduler are $pendingTasks$ and $currentlyBeingProcessed$.
The $pendingTasks$ is the array of pending tasks and $currentlyBeingProcessed$ is a boolean array indexed by the tasks where $currentlyBeingProcessed[\task]$ denotes whether task $\task$ is currently being processed by some resource.
Once a task is assigned to some resource and the resource starts to process it, the task cannot be moved to another resource.
The global variables are kept in the memory for the whole running time of the scheduling system.

The scheduling policy receives two input arguments (in addition to the global variables).
The first one, denoted as $t$, is the time when the event leading to rescheduling occurred.
The second one, denoted as $\probsol$, is the current solution to the scheduling problem.

The scheduling policy returns a schedule of the pending tasks which is an array $sch$ indexed by the resources.
Each element $sch[\resource]$ of this array is an another array which defines the order of the assigned tasks on this resource.

The policy starts by performing the initialization of the $earliestStartTime$ and $sch$ arrays.
The $earliestStartTime$ array represents the earliest start time of the tasks on each resource with respect to the previous assignments.
Then, in the loop at line \ref{src:edd_mct/adding_assigned_tasks_to_schedule}, the currently processed tasks on the resources are added to the start of the schedule of each resource.
The loop at line \ref{src:edd_mct/not_scheduling_in_past} ensures that the rest of the pending tasks cannot be assigned in the past.
The next part sorts the tasks by their requested due dates, i.e. this is the selection policy.
The last loop at line \ref{src:edd_mct/allocation} finds a resource for each pending task such that it can finish that particular task at the earliest with respect to the previously assigned tasks, i.e. this is the assignment policy.

\begin{algorithm}[t]
  \SetKwFunction{KwFn}{EDDMCT}
  \SetKw{Fn}{Function}
  \Fn{}
  \KwFn{$t, pendingTasks, \probsol, currentlyBeingProcessed$}
  \SetKwBlock{BeginAlgorithm}{}{}
  \BeginAlgorithm{
    \tcc{Initialization}
    \ForEach{$\resource \in \resourceset$}{
      $earliestStartTime[\resource] \gets 0$\;
      $sch[\resource] \gets$  empty array\;
    }
    \tcc{Add the already assigned tasks to the schedule}
    \ForEach{$\task \in pendingTasks$}{
      \label{src:edd_mct/adding_assigned_tasks_to_schedule}
      \If{$currentlyBeingProcessed[\task]$} {
        $earliestStartTime[\rallocsol*] \gets \ecompfn\left(\task, \rallocsol*, \allocsol*, 1\right)$\;
        $sch[\rallocsol*][end + 1] \gets \task$\;
      }
    }
    \ForEach{$\resource \in \resourceset$}{
      \label{src:edd_mct/not_scheduling_in_past}
      $earliestStartTime[\resource] \gets \max(earliestStartTime[\resource], t)$\;
    }
    $sorted \gets$ array of sorted tasks from $pendingTasks$ in ascending order by (i) $\reqdue*$, if equal then (ii) $\enptfn(\instance*, \qfnsol*)$, if equal then (iii) $\arrival*$, if equal then (iv) randomly\;
    \tcc{Assignment policy}
    \ForEach{$k = 1, \dots, len(sorted)$}{
      \label{src:edd_mct/allocation}
      $\task \gets sorted[k]$\;
      \If{$currentlyBeingProcessed[i] = false$}{
        $\rallocsol* \gets \argmin_{\resource \in \resourceset} \ecompfn\left(\task, \resource, earliestStartTime[\resource], 1\right)$\;
        $\allocsol* \gets earliestStartTime[\rallocsol*]$\;
        $earliestStartTime[\rallocsol*] \gets \ecompfn\left(\task, \rallocsol*, \allocsol*, 1\right)$\;
        \If{$len(sch[\rallocsol*]) = 0$}{
          \tcc{The resource starts processing the task immediately}
          $currentlyBeingProcessed[i] \gets true$\;
        }
        $sch[\rallocsol*][end + 1] \gets \task$\;
      }
    }
    \Return{$sch$}\;
  }
  \caption{EDDMCT (Earliest Due Date + Minimal Completion Time) scheduling policy}
  \label{alg:alg/edd_mct}
\end{algorithm}

Since the environment in which the scheduling system operates is not static, the scheduling system needs to adapt to the changes in this environment.
These changes are propagated to the scheduling system as \define{events}.
When the environment changes, the scheduling system receives a corresponding event and runs the scheduling policy to adapt to the new state of the environment.
The following events are recognized by the scheduling system (if an additional procedure needs to be performed before the scheduling policy is run, the description of that procedure is also provided):
\begin{itemize}
  \item New task $\task$ was received by the scheduling system: task $\task$ is added to the array of pending tasks, i.e.

    {
      \LinesNotNumbered
      \RestyleAlgo{plain}
      \begin{algorithm}[H]
        $currentlyBeingProcessed[\task] \gets false$\;
        $pendingTasks \gets pendingTasks \cup \{\task\}$\;
      \end{algorithm}
    }

  \item The solution for some task $\task$ was received by the scheduling system: task $\task$ is removed from the array of pending tasks, i.e. 

    {
      \LinesNotNumbered
      \RestyleAlgo{plain}
      \begin{algorithm}[H]
        $currentlyBeingProcessed[\task] \gets false$\;
        $pendingTasks \gets pendingTasks \setminus \{\task\}$\;
      \end{algorithm}
    }
\end{itemize}

The complexity of the presented policy is $\mathcal{O}\left(\npendtasks \log{\npendtasks} + \npendtasks \nresources \right)$.
The term $\npendtasks \log{\npendtasks}$ is due to sorting of the pending tasks.
The term $\npendtasks \nresources$ is added because it is needed to iterate over all the resources to find a resource which completes the given task at the earliest.

\subsection{Quality control}
\label{sec:alg/quality}
We propose two quality control algorithms: (i) \define{bisection control} and (ii) \define{individual control}.
Both algorithms are able to handle the overload situations.
However, they have different behavior and assumptions.

\subsubsection{Bisection control}
The bisection control algorithm is described in Algorithm \ref{alg:alg/quality}.
The algorithm tries to find one global quality $\q$ for all the tasks over all the resources such that the average normalized lateness of the tasks in the current schedule is close to 0.
Equivalently, the algorithm tries to find a root of the average normalized lateness function for the fixed assignment and order of the tasks defined in $sch$, i.e. output of Algorithm \ref{alg:alg/edd_mct}.
For efficiency and ease of implementation, a well-known \define{bisection method} is used as the root-finding method.
A bisection method iteratively divides the quality interval $\qset$ into two halves.
The search continues only in the half where the root lies.
Algorithm \ref{alg:alg/quality} also uses Algorithm \ref{alg:alg/sch_recomp} which recomputes the duration of the tasks in the current schedule for the given fixed quality $\q$ and returns the average normalized lateness of the tasks in the current schedule. 

In addition to the already introduced input arguments for the scheduling policy, the control algorithm has three new input arguments.
The first one, denoted as $sch$, is the schedule created by Algorithm \ref{alg:alg/edd_mct}.
The second one, denoted as $maxIters$, is the maximum number of iterations of the control algorithm.
The third one, denoted as $\minq$, is the minimum quality of the solutions.
Both arguments $maxIters$ and $\minq$ are set by the administrator of the scheduling system depending on the desired behavior.
The $maxIters$ argument controls the performance of the control algorithm: the higher number of iterations will result in a more accurate quality, but the running time of the control algorithm will be higher.
Based on our preliminary experiments, for the majority of the cases, the sufficient number of iterations is 30.

The complexity of Algorithm \ref{alg:alg/quality} is $\mathcal{O}\left(maxIters \cdot \left(\nresources + \npendtasks\right)\right)$, where $\nresources + \npendtasks$ is due to schedule recomputation algorithm.

\begin{algorithm}[t]
  \SetKwFunction{KwFn}{BisectionControl}
  \SetKw{Fn}{Function}
  \Fn{}
  \KwFn{$t, \probsol, sch, maxIters, \minq$}
  \SetKwBlock{BeginAlgorithm}{}{}
  \BeginAlgorithm{
    $averageNormalizedLateness \gets ScheduleRecomputation(t, sch, 1)$\;
    \lIf{$averageNormalizedLateness \le 0$}{
      \Return
    }
    $averageNormalizedLateness \gets ScheduleRecomputation(t, sch, \minq)$\;
    \lIf{$averageNormalizedLateness \ge 0$}{
      \Return
    }
    $iter \gets 0$; $lb \gets \minq$; $ub \gets 1$; $\q \gets 1$\;
    $averageNormalizedLateness \gets 1$\;
    \While{$iters < maxIters$}{
      \eIf{$averageNormalizedLateness > 0$}{
        $ub \gets \q$\;
      }{
        $lb \gets \q$\;
      }
      $\q \gets \frac{lb + ub}{2}$\;
      $averageNormalizedLateness \gets ScheduleRecomputation(t, sch, \q)$\;
      $iter \gets iter + 1$\;
    }
  }
  \caption{Bisection control algorithm}
  \label{alg:alg/quality}
\end{algorithm}

\begin{algorithm}[t]
  \SetKwFunction{KwFn}{ScheduleRecomputation}
  \SetKw{Fn}{Function}
  \Fn{}
  \KwFn{$t, sch, \q$}
  \SetKwBlock{BeginAlgorithm}{}{}
  \BeginAlgorithm{
    $sumNormalizedLateness \gets 0$; $numTasks \gets 0$\;
    \ForEach{$\resource \in \resourceset$}{
      $earliestStartTime[\resource] \gets 0$\;
      \ForEach{$k = 1, \dots, len(sch[\resource])$}{
        $\task \gets sch[\resource][k]$\;
        $\qfnsol*(t) \gets \q$\;
        \If{$k \not= 1$}{
          $\allocsol* \gets earliestStartTime[\resource]$\;
        }
        $earliestStartTime[\resource] \gets \ecompfn\left(\task, \resource,\allocsol*, \q\right)$\;
        \If{$k = 1$}{
            $earliestStartTime[\resource] \gets \max\{t, earliestStartTime[\resource]\}$\;
        }
        $sumNormalizedLateness \gets sumNormalizedLateness + \frac{earliestStartTime[\resource] - \reqdue*}{\reqresp*}$\;
      }
      $numTasks \gets numTasks + len(sch[\resource])$\;
    }
    \eIf{$numTasks = 0$}{
      \Return 0\;
    }{
      \Return{$\frac{sumNormalizedLateness}{numTasks}$}\;
    }
  }
  \caption{Schedule recomputation}
  \label{alg:alg/sch_recomp}
\end{algorithm}

\subsubsection{Individual control}
\label{sec:alg/quality/ind}
The individual control algorithm is described in Algorithm \ref{alg:alg/individual}.
The quality is computed for each task and resource independently, and the algorithm assumes that the normalized processing time of each task $\task$ can be approximated by a linear function $\qfnsol*(t)\cdot \enptfn(\instance*, 1)$.
To illustrate the concept behind the algorithm, we can assume that: (i) all tasks from $\taskset$ are processed on one resource $\resource$, (ii) the order of assignments is given by the value of $\task \in \taskset$, and (iii) resource $\resource$ has no idle times between processing each task.
First, expand Equation \eqref{eq:nalat} for the average normalized lateness using the estimated completion time \eqref{eq:ecompfn}
\begin{equation}
  \label{eq:nalat_expand}
  \begin{aligned}
    \frac{1}{\ntasks}\sum_{\task \in \taskset} \frac{\ecompfn(\task, \resource,
    \allocsol[\task], \qfnsol[\task](t)) - \reqdue*}{\reqresp*} &=
    \frac{1}{\ntasks}\sum_{\task \in \taskset} \frac{\allocsol[\task] +
        \qfnsol*(t) \frac{\enptfn(\instance*,
    1)}{\speed*} - \reqdue*}{\reqresp*} \\
    &=
    \frac{1}{\ntasks}\sum_{\task \in \taskset} \frac{\allocsol[1] + \left(\sum_{k =
            1}^{\task} \qfnsol[k](t) \frac{\enptfn(\instance[k],
    1)}{\speed*}\right) - \reqdue*}{\reqresp*} \\
    &=
    \left(
    \sum_{\task \in \taskset} \sum_{k = 1}^{\task} \qfnsol[k](t) \frac{\enptfn(\instance[k], 1)}{\speed* \cdot \ntasks} \frac{1}{\reqresp[\task]}
    \right)
    +
    \left(
    \sum_{\task \in \taskset} \frac{\allocsol[1] - \reqdue*}{\ntasks \cdot \reqresp*}
    \right)
    \,.
  \end{aligned}
\end{equation}

The second term of Equation \eqref{eq:nalat_expand} on the right-hand side is a constant, i.e. its value does not depend on the values $\qfnsol*(t)$.
This term is denoted as $cons$ in Algorithm \ref{alg:alg/individual}.
To further simplify the first term, the sums for each $\task$ are expanded and the multipliers of each $\qfnsol*(t)$ are collected.
By doing so, \define{a weight} of each task is obtained
\begin{equation}
  \weight* = \frac{\enptfn(\instance*,
  1)}{\speed* \cdot \ntasks} \sum_{k = \task}^{\ntasks}
  \frac{1}{\reqresp[k]}\,,
\end{equation}
with which the average normalized lateness from Equation \eqref{eq:nalat_expand} can be rewritten as
\begin{equation}
  \label{eq:ind_control_final_nalat}
  \left(
  \sum_{\task \in \taskset} \qfnsol*(t) \cdot \weight*
  \right)
  +
  \left(
  \sum_{\task \in \taskset} \frac{\allocsol[1] - \reqdue*}{\ntasks \cdot \reqresp*}
  \right)
  \,.
\end{equation}
From this equation, it can be seen that the tasks that contribute the most to the average normalized lateness are those with the highest weights.
The idea of the individual control algorithm is to compress greedily the tasks with the highest weights until the average normalized lateness is close to zero.

The algorithm starts by computing the weight of each task and the constant part $cons$ (see line \ref{src:individual/cons}).
Then, the tasks are sorted non-increasingly by their weights (see line \ref{src:individual/sorting_by_weight}) and the algorithm proceeds by greedy compression of the tasks with the highest weights (see line \ref{src:individual/compressing_cycle}).
The compression is at line \ref{src:individual/compression} where the qualities for all tasks except $\task$ are fixed and the algorithm is trying to find the root of Equation \eqref{eq:ind_control_final_nalat}.
The last part of the algorithm recomputes the start time of each task using the computed qualities (see line \ref{src:individual/start_time_recomputation}).

There are two complications which need to be considered when implementing this idea.
The first complication is that if some task is currently being processed by a resource, generally it cannot be compressed to $\minq$ because the solution found until time $t$ can already be of better quality than $\minq$, i.e. the quality of the current solution is a new lower bound for the quality compression.
Therefore, the quality of the current solution must be reflected in the algorithm when compressing the first task (see line \ref{src:individual/quality_not_lost}).
The second complication arises when the requested quality for the first task is set to one and the solution is not received within the estimated completion time, e.g. the solution is being sent from a resource to the scheduling system.
In such a case, there is an idle time between the estimated completion time of the first task and time $t$ called the \define{phantom time}, which needs to be considered when computing the average normalized lateness (see lines \ref{src:individual/phantom_1} and \ref{src:individual/phantom_2}).

\begin{algorithm}[H]
  \SetKwFunction{KwFn}{IndividualControl}
  \SetKw{Fn}{Function}
  \Fn{}
  \KwFn{$t, \probsol, sch, \resource, \minq$}
  \SetKwBlock{BeginAlgorithm}{}{}
  \BeginAlgorithm{
    \lIf{$len(sch[\resource])=0$}{\Return}
    $usePhantom \gets false$; $phantomTime \gets 0$\;
   \If{$t \ge \ecompfn(sch[\resource][1], \resource, \allocsol[sch[\resource][1]], 1)$}{
     $usePhantom \gets true$\;
     $phantomTime \gets t - \ecompfn(sch[\resource][1], \resource, \allocsol[sch[\resource][1]], 1)$\;
   }
   \tcc{The constant part of the average normalized lateness}
   $cons \gets 0$\;
   \tcc{Compute the weights and the constant part}
   \ForEach{$k = 1, \dots,len(sch[\resource])$}{
     \label{src:individual/cons}
     $\task \gets sch[j][k]$\;
     $\qfnsol*(t) \gets 1$\;
     $w[i] = \frac{\enptfn(\instance*, 1)}{\speed*\cdot len(sch[\resource])} \sum_{l = k}^{len(sch[j])} \frac{1}{\reqresp[sch[\resource][l]]}$\;
     $cons \gets cons + \frac{\allocsol[sch[\resource][1]] + phantomTime- \reqdue*}{len(sch[\resource])\cdot\reqresp*}$\;
     \label{src:individual/phantom_1}
   }
   $sorted \gets$ array of tasks from $sch[\resource]$ sorted decreasingly by $w$\;
   \label{src:individual/sorting_by_weight}
   \tcc{Compress the tasks in order given by the $sorted$ array}
   \ForEach{$k = 1,\dots,len(sorted)$}{
       \label{src:individual/compressing_cycle}
       \If{$\left(\sum_{l = 1}^{len(sch[\resource])}\qfnsol[sch[\resource][l]](t) \cdot w[sch[\resource][l]] \right) + cons$ \text{\normalfont is negative or is near to zero}}{
         \tcc{Stop if the average normalized lateness is}
         \tcc{negative or is near to zero}
         \textbf{break}\;
       }
       $\task \gets sorted[k]$\;
       \If{$\task = sch[\resource][1]\,\wedge\,usePhantom$}{
           \label{src:individual/phantom_2}
           \tcc{The first task cannot be compressed further}
           \textbf{continue}\;
       }
       \eIf{$\task = sch[\resource][1]$}{
         \tcc{The current progress of the task should not be ``lost''}
         $qualityLowerBound \gets \max\left\{\minq, \frac{(t - \allocsol*)\speed*}{\enptfn(\instance*, 1)}\right\}$\;
         \label{src:individual/quality_not_lost}
       }{
         $qualityLowerBound \gets \minq$\;
       }
       $\qfnsol*(t) \gets \max\left\{qualityLowerBound, \frac{-cons - \sum_{l = 1, l\not=i}^{len(sch[\resource])} \qfnsol[sch[\resource][l]](t) \cdot w[sch[\resource][l]]}{w[\task]}\right\}$\;
       \label{src:individual/compression}
    }
    $earliestStartTime \gets 0$\;
    \tcc{Computation of the start times using the found qualities}
    \ForEach{$k = 1, \dots, len(sch[\resource])$}{
      \label{src:individual/start_time_recomputation}
      $\task \gets sch[\resource][k]$\;
      \If{$k \not= 1$}{
        $\allocsol* \gets earliestStartTime$\;
      }
      $earliestStartTime \gets \ecompfn\left(\task, \resource, \allocsol*, \qfnsol*(t)\right)$\;
      \If{$k = 1$}{
        $earliestStartTime \gets \max\{t, earliestStartTime\}$\;
      }
    }
  }
  \caption{Individual control algorithm}
  \label{alg:alg/individual}
\end{algorithm}

In addition to the already introduced input arguments for the bisection control, the independent control algorithm has one new input argument $\resource$ which denotes the resource for which the qualities are currently computed.

The complexity of the algorithm is $\mathcal{O}\left(\ntasks[\resource] \log{\ntasks[\resource]}\right)$ because the complexity is dominated by the sorting of the tasks.
From Theorem \ref{thm:alg/individual/total_comp}, the total complexity of computing Algorithm \ref{alg:alg/individual} for all resources is $\mathcal{O}\left(\npendtasks \log{\npendtasks}\right)$.
\begin{thm}
  \label{thm:alg/individual/total_comp}
  The total worst case complexity of computing Algorithm
  \ref{alg:alg/individual} for all resources is $\mathcal{O}\left(\npendtasks
  \log{\npendtasks}\right)$.
\end{thm}
\begin{pf}
  See \ref{sec:app/comp}.
\end{pf}

%% file: estimation.tex
\section{Estimation of the normalized processing time functions}
\label{sec:estim}
Since normalized processing time function $\nptfn$ is not known \textit{a priori}, the scheduling system needs to estimate it\footnote{For simplicity, we assumed in this work that the instances are only of one optimization problem, i.e. personnel rostering. However, it is possible that the scheduling system would have to process more optimization problems. In such a case, the processing time functions would be indexed by the problems and the estimation would be performed for each problem individually.}.
This estimation can be used by the scheduler instead of the worst-case estimates to find better assignments.

\begin{figure}[H]
  \centering
  \includegraphics[scale=1]{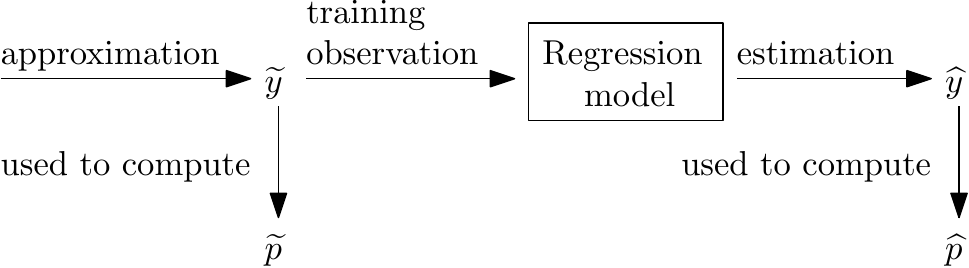}
  \caption{The flow of the normalized processing time function estimation.}
  \label{fig:alg/proc_flow}
\end{figure}

Estimation is based on the assumption that $\nptfn$, for the given instance, can be approximated by a \define{piecewise linear function} which is parameterized by \define{approximated parameters} $\afitparams$.
Each parameter $\afitparam_k$ represents the normalized processing time of the endpoint of $k$-th segment.
An endpoint is a point where two neighboring segments of a piecewise linear function meet.
The approximation is made due to performance reasons - a function approximation with a few parameters can be more efficiently estimated than the whole $\nptfn$.
However, since $\afitparams*$ are not known \textit{a priori} (because $\nptfn$ is also not known \textit{a priori}), the question remains how to estimate them when task $\task$ arrives in the scheduling system.
To solve this problem, a \define{regression analysis} is used.
Using a set of previously collected parameters $\afitparams$, a \define{regression model} is \define{trained}.
That regression model is then able to estimate $\afitparams*$ for the given task $\task$, where the \define{estimated parameters} for task $\task$ are denoted as $\efitparams*$.
Figure \ref{fig:alg/proc_flow} summarizes the flow of the normalized processing time estimation.

The training of the regression model can be performed:
\begin{itemize}
  \item \define{offline}, i.e. before the scheduling system is used in production.

  \item \define{online}, i.e. after enough observations are collected at runtime.

  \item using a combined approach.
\end{itemize}
In the combined approach, the estimator starts with an initial regression model which is iteratively refined using information acquired online.
The offline model can be built by sampling the observation space, i.e. the space of tuples representing the features and the approximated parameters of the instances.
However, this method assumes that we are provided with such samples.
The online model does not need such samples beforehand.
However, its disadvantage is that until enough training observations are collected, the estimator must use the default parameters such as the worst case normalized processing time.
All approaches also fail to provide satisfactory estimations if the observation space is insufficiently sampled, e.g. observations used for training the regression model are sampled from a different region of the observation space than the observations to estimate.

\subsection{Collecting training observations}
\label{sec:estim/collect}
\begin{figure}[H]
  \centering
  \includegraphics[scale=0.45]{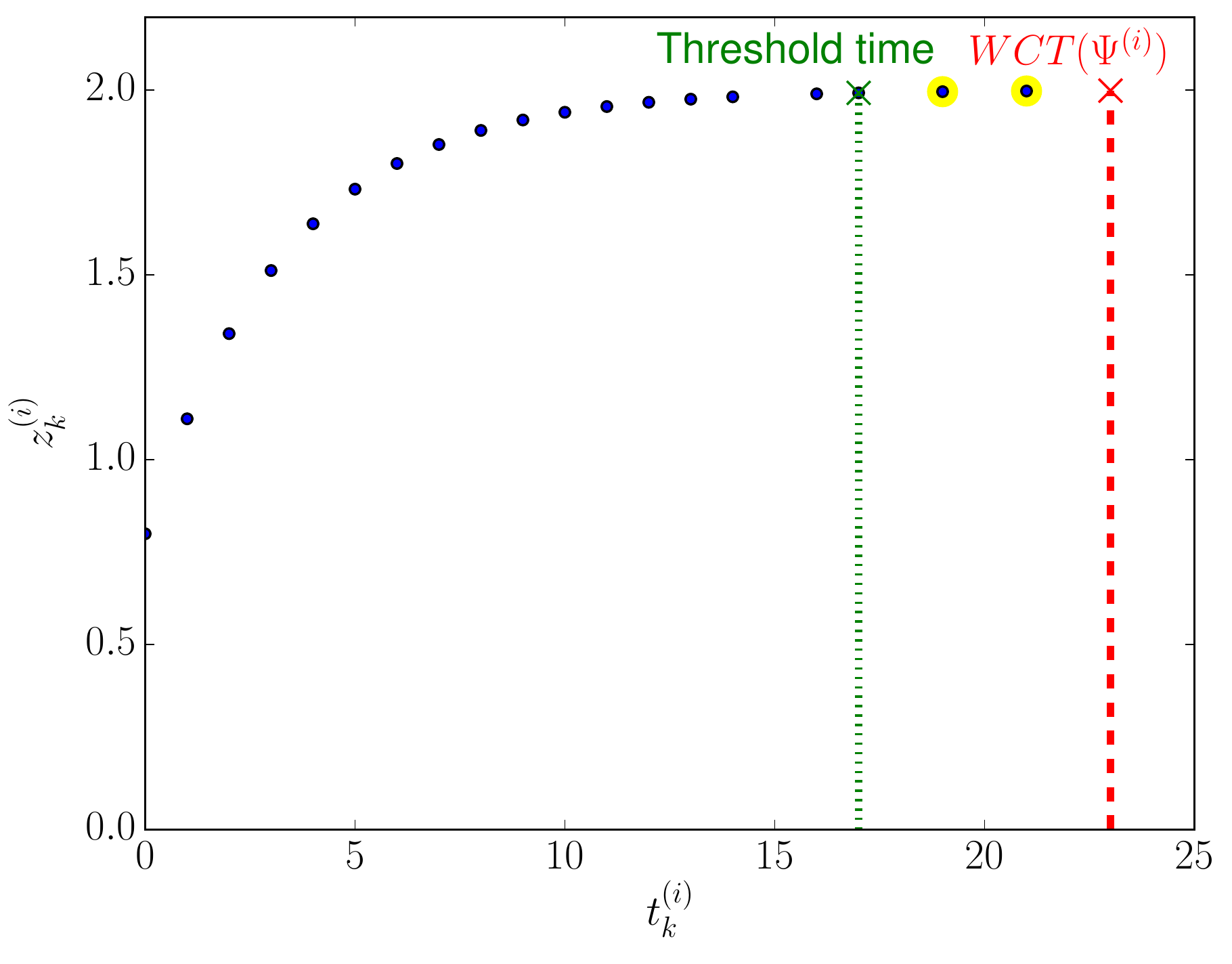}
  \caption{
    An example of a measured objective curve with highlighted worst case processing time ($\nwcttimefn$) and the threshold time denoting the processing time for solution quality of 1.
    The trimmed points are highlighted by larger yellow circles.
  }
  \label{fig:alg/criterion_curve}
\end{figure}

This Subsection describes how the training observations are collected and processed.
When the online learning is employed, the steps are performed during the runtime of the scheduling system.
If the offline learning is used, the steps are performed on the sampled observations before the scheduling system is used in production.
A new training observation from task $\task$ is created in the following steps:
\begin{enumerate}
  \item The scheduling system assigns task $\task$ to some resource $\resource$ with normalized processing time equal to the normalized worst case processing time of that task.
    The normalized worst case processing time represents some fixed time after which the algorithm is stopped even though the solution could still be improved.
    However, it is our assumption that the improvements after the normalized worst case processing time are insignificant. 

  \item The algorithm solving the task must record how the value of the objective for the best-known solution evolves through time.
    This progress is recorded in a \define{objective curve} $\critcurve*$ of task $\task$ which is a sequence of tuples
    $
      \left( \left(\critcurvet*_1, \critcurvev*_1\right),
                            \left(\critcurvet*_2, \critcurvev*_2\right),
                            \ldots,
                            \left(\critcurvet*_{h_i}, \critcurvev*_{h_i}\right)
                    \right)\,,
    $
    where $\critcurvev*_k$ is a objective value of the best known solution until time $\critcurvet*_k$ and $h_i$ is a number of collected tuples.
    Both $\critcurvev*_k$ and $\critcurvet*_k$ are assumed to be increasing sequences, i.e. it is assumed that the objective function is maximized.
    The objective function can also be minimized, but then the algorithm solving the task is responsible for transforming the decreasing objective curve to an increasing one.
    
  \item When the task is finished, resource $\resource$ sends the solution and objective curve $\critcurve*$ back to the central scheduler which sends the objective curve to the estimator.

  \item The estimator rescales times $\critcurvet*_k$ in the objective curve by the estimated speed of resource $\resource$ to the normalized processing time $\critcurvet*_k \speedfn*\,.$

  \item The objective curve is trimmed to retain only the times $\critcurvet*_k \le \nwcttimefn(\instance*)\,.$

  \item The objective curve is trimmed to retain only the part where the approximated slope of the objective value is higher than some threshold, i.e. this step performs the ``tail cutting''.
    The slope for each time $\critcurvet*_k$ is computed as 
      \begin{equation}
        \label{eq:slope_trim}
        slope_k = \frac{\critcurvev*_{k+1}
        - \critcurvev*_{k}}{\critcurvet*_{k+1} - \critcurvet*_{k}}
      \end{equation}

      and the threshold is computed as $0.05\cdot \max_k slope_k\,$.
      This is the last step where the objective curve is trimmed; let us denote the index of the last remaining tuple from the objective curve as $k_{max}$.

      The reason for this step is that even if the normalized worst case processing time represents the stopping time of the algorithm, it does not mean that the solution of the maximum quality cannot be found much earlier or that all solutions found after some threshold time are significantly better than the solution found by the threshold time.
      Consider Figure \ref{fig:alg/criterion_curve} which illustrates this idea.
      The threshold time represents the time after which all measurement are trimmed because the improvements of the solution are not significant.
      The best-known solution until this threshold time is then declared as the solution of the maximum quality.

  \item objective values $\critcurvev*_k$ are linearly scaled to $[0,1]$ using min-max normalization.
    The normalized values are denoted as $\critcurveq*_k$.

  \item The curve from the previous step represents the progress of the normalized processing time on the solution quality.
    The number of tuples in this curve can be large and, therefore, a compact representation by the approximated curve is used.
    For our purposes, a piecewise linear function with ten segments is used.
    The endpoints of the qualities in this piecewise linear function are fixed to $0.1, 0.2, \dots, 0.9,1$.
    To acquire the approximated parameters $\afitparams*$, a curve from the previous step is sampled along those fixed endpoints, e.g. $\afitparam*_3$ represents the normalized processing time needed to find the solution of quality 0.3.
    A linear interpolation between the neighboring qualities is used if the value of the given fixed quality does not exist in the curve.

  \item A new training observation $\left(\features*, \afitparams*\right)$ is added to the database of the training observations, where $\features* = \featurefn(\instance*)$ is the feature vector of task $\task$.
\end{enumerate}

\subsection{Regression model training and estimation of unseen tasks}
When a sufficient number of training observations is collected, a regression model is trained.
The regression model $\modelfn$ maps the feature vectors $\features* = \featurefn(\instance*)$ to the estimated parameters, i.e. $\efitparams* = \modelfn(\features*)$.
The found parameters are then used to construct the estimated normalized processing time function.
Since a piecewise linear function is used for approximating the objective curve, the same function type is also used for the estimation.
Therefore, the coordinates of the endpoints in the estimated normalized processing time function are $(0.1 \cdot k, \efitparam*_l)$, where $k = 1, \dots, 10$.

Since most regression methods learn only one output, an independent regression model is trained for each quality endpoint $0.1 \cdot k$ and outputs of those models are then combined into one function.
However, this approach may result in a non-monotonic function; this can be corrected by taking a running maximum of the processing time values.

\subsubsection{Estimation methods}
\label{sec:alg/estim/methods}
In the previous Subsection a procedure for collecting training observations using an approximation by a piecewise linear function is described.
This approach is called \define{full estimation}.
For the experiments in Section \ref{sec:exp}, the following alternative estimation methods are also considered:
\begin{enumerate}
  \item \define{measured processing time}: No estimation is employed, the scheduling system has full knowledge of the processing time function (the function is acquired by running all instances before the scheduling system is run and measuring their processing time).
    Since in most cases this knowledge is unavailable, such scenario is unrealistic in practice.
    However, in experiments such scenario allows us to determine how the performance of the scheduling system differs from a more realistic scenario where the estimation is used.

  \item \define{linear estimation}: Instead of using a piecewise linear function to approximate the measured curve, a simple linear function can be used.
    In such case, only one point needs to be estimated to obtain the estimated processing time function.
\end{enumerate}

\begin{figure}[h]
  \centering
  \includegraphics[scale=0.5]{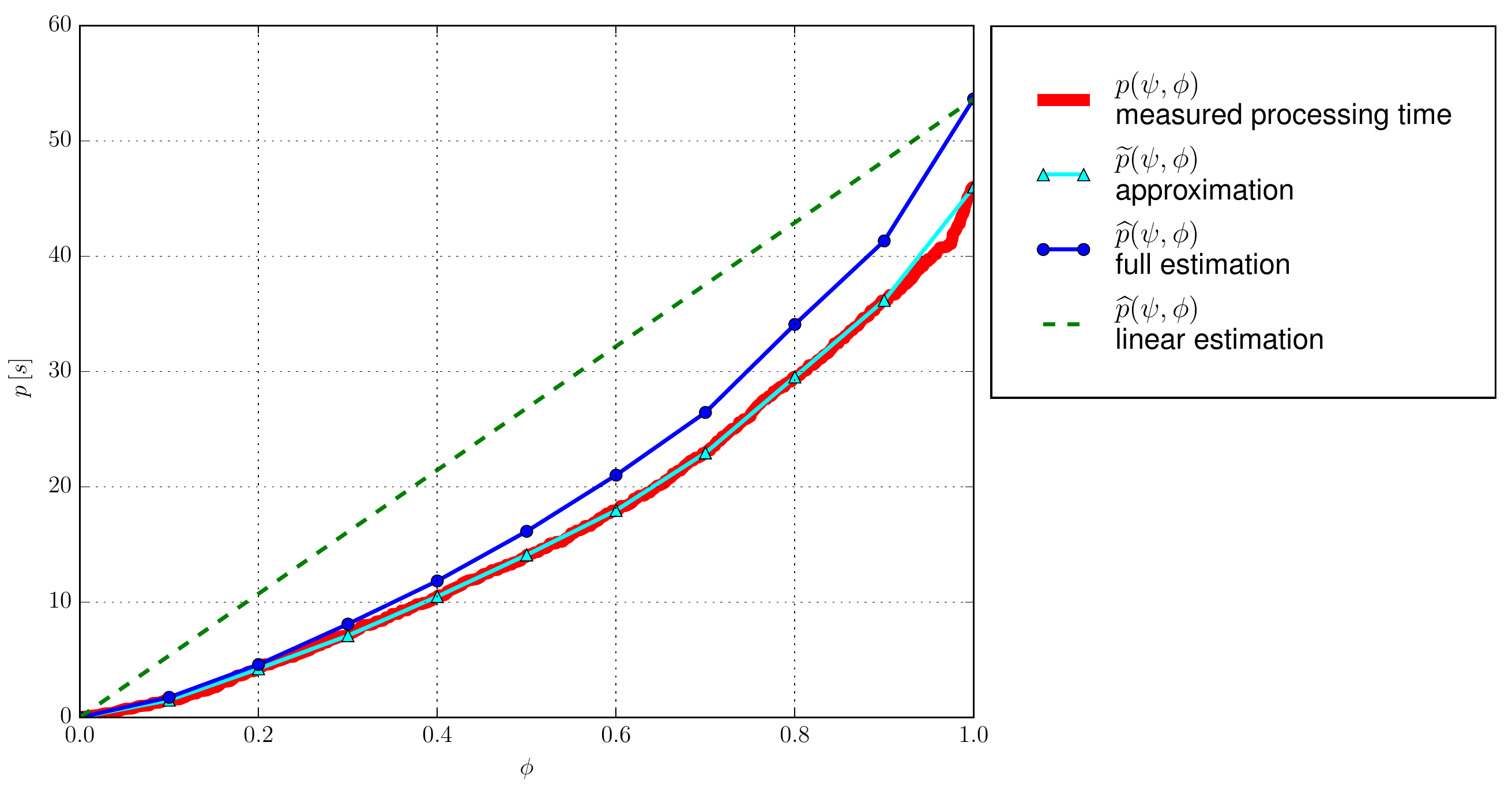}
  \caption{
    An example of the measured processing time (drawn in red, solid line), its approximation (drawn in cyan, solid line with triangles), its full estimation (drawn in blue, solid line with points) and its linear estimation (drawn in green, dashed line).
    The triangles and points in approximation and full estimation, respectively, represent the endpoints of the piecewise linear function.
    The estimation is found using the $k$-nn method.
  }
  \label{fig:alg/estim/methods/objective_curves}
\end{figure}

All approaches are illustrated in Figure \ref{fig:alg/estim/methods/objective_curves}.

%% file: experiments.tex
\section{Experiments}
\label{sec:exp}
In this section, it is verified that our proposed scheduling system can keep the average normalized lateness near 0.
It is shown that if the quality control is disabled, the lateness could be 60 times larger than the requested lateness which is unacceptable.

First, the experimental results for the offline estimation of the processing time of real-world instances from a domain of personnel rostering are provided.
Then, the instances from the previous experiment and their estimations are used for the experiments with the scheduling system.
The computer, on which the scheduling system run, has an Intel Core i7-3520M @ 2.90GHz processor and 8 GB of RAM.

\subsection{Processing time estimation}
\subsubsection{Experimental setup}
As a task type for experiments, an \define{automatic roster design} from the domain of personnel rostering is used.
Initially, 13 features of the rostering instances were considered (some similar to ones described in \cite{messelis2013a}), but after performing the sequential forward selection, 8 features were obtained from which the following features are the most influential: the number of employees, the number of days, the number of required shifts to assign and roster size (the number of employees multiplied by the number of days).
Since each feature has a different scale of values, the features are normalized using \define{z-score normalization} so that each feature has the same weight.

To generate the dataset, 500 observations were generated which differed in number of the employees (5-30, sampled uniformly), number of days in a roster (7, 14 and 31, sampled uniformly), weekly workload of the employees (20 and 40 hours, sampled uniformly) and workload coverage (ratio between the number of required shifts to the number of shifts computed from the workload of the employees).
All observations had three types of shifts to assign: early, late and night (each shift was 8 hours long).
As a \define{training set}, 300 observations were randomly selected from the original set of 500 observations; the rest was used as a \define{testing set}.
The training set was used for feature selection, parameter tuning of the regression methods and training the estimation model.
The testing set was used to determine the performance of the learned model on unknown data.

Two methods were considered for the estimation: (i) \define{k-nearest neighbors} (abbreviated as $k$-nn) and (ii) \define{regression trees}.
For the $k$-nearest neighbors method, the Euclidean distance was chosen as a distance metric and the number of neighbors was set to 7.
For the regression tree method, the minimum number of observations in the branch nodes was set to 10 and the minimum number of observations in the leaf nodes was set to 1.

\subsubsection{Results}
\begin{figure}[H]
  \centering
  \includegraphics[scale=0.5]{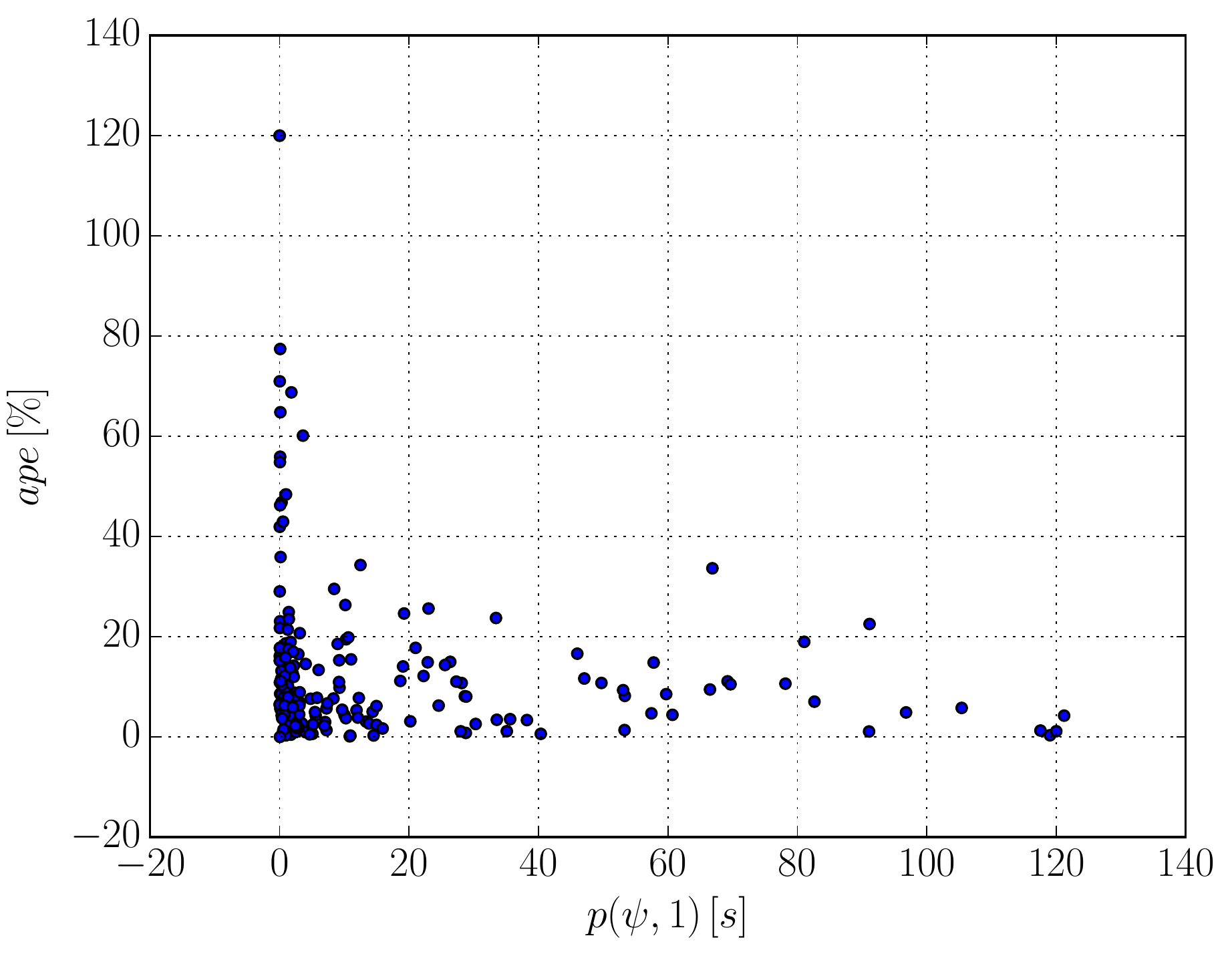}
  \caption{
    The relationship between the absolute percentage error and the maximum normalized processing time.
    Each point represents one observation estimated using the $k$-nn method.
    The total number of observations is 200, i.e. the whole testing set.
  }
  \label{fig:exp/est/aet_vs_ape}
\end{figure}

The results of the estimation experiment are reported in Table \ref{tab:exp/est/ape} and in Figure \ref{fig:exp/est/aet_vs_ape}.
As an error metric, an \define{absolute percentage error} is used and is defined as
\begin{equation}
  \ape(\nptfn, \enptfn, \q, \instance) = \abs{\frac{\enptfn(\instance, \q) - \nptfn(\instance, \q)}{\nptfn(\instance, \q)}} \cdot 100\%\,.
\end{equation}
The absolute percentage error for the given solution quality $\q$ and instance $\instance$ is an error ratio between the normalized processing time of instance $\instance$ for quality $\q$ and the estimated normalized processing time of instance $\instance$ for quality $\q$.
The qualities in Table \ref{tab:exp/est/ape} correspond to the endpoint coordinates of a piecewise linear function.
Figure \ref{fig:exp/est/aet_vs_ape} shows the absolute percentage error of each testing observation for maximum quality.

Notice that for higher qualities the absolute percentage error of the majority of observations ($0.75\,\%$) is around $20\,\%$ for both methods.
For this quantile, the $k$-nn method gives better estimations with the exception of the $0.2$ quality.
The higher estimation error is achieved when the processing time is low, as can be seen in Figure \ref{fig:exp/est/aet_vs_ape}.
This is not a considerable issue, because the absolute error, i.e. the absolute value of the difference between $\fitparam$ and $\efitparam$, of these short instances is small.
Therefore, the overall effect on the scheduling system is also small if the system processes both short and long tasks.
To overcome the problem of underestimation of short tasks, the minimum processing time can be set to some fixed lower bound, e.g. $1\,s$.
Because the $k$-nn method achieves lower estimation error on the majority of the observations, the $k$-nn method is used in the following experiments of the scheduling system.

\begin{table}[H]
    \centering
\begin{tabular}{|cccccccc|}
\hline
\multicolumn{2}{|c}{} & \multicolumn{6}{c|}{Absolute percentage error as quantiles [\%]} \\ 
\hline
Quality & Method & 0.05 & 0.25 & 0.5 & 0.75 & 0.95 & 1.00 \\ 
\hline
\multirow{2}{*}{0.1}  &\cellcolor{gray!25} regression tree  & \cellcolor{gray!25} 2.0  &\cellcolor{gray!25} 7.9  &\cellcolor{gray!25} 19.8  &\cellcolor{gray!25} 34.6  &\cellcolor{gray!25} 78.3  &\cellcolor{gray!25} \textbf{172.0} \\ \cline{2-8}
 & knn  & \textbf{1.4}  & \textbf{7.2}  & \textbf{18.1}  & \textbf{32.1}  & \textbf{67.3}  & 186.3 \\ 
\hline
\multirow{2}{*}{0.2}  &\cellcolor{gray!25} regression tree  &\cellcolor{gray!25} \textbf{1.2}  &\cellcolor{gray!25} \textbf{6.2}  &\cellcolor{gray!25} 13.7  &\cellcolor{gray!25} \textbf{24.5}  &\cellcolor{gray!25} \textbf{49.1}  &\cellcolor{gray!25} \textbf{146.1} \\ \cline{2-8}
 & knn  & 1.3  & 6.6  & \textbf{13.1}  & 28.6  & 60.5  & 170.2 \\ 
\hline
\multirow{2}{*}{0.3}  &\cellcolor{gray!25} regression tree  &\cellcolor{gray!25} 1.4  &\cellcolor{gray!25} 7.1  &\cellcolor{gray!25} 14.8  &\cellcolor{gray!25} 25.6  &\cellcolor{gray!25} \textbf{52.7}  &\cellcolor{gray!25} 140.3 \\ \cline{2-8}
 & knn  & \textbf{1.0}  & \textbf{6.2}  & \textbf{13.1}  & \textbf{24.8}  & 63.1  & \textbf{132.4} \\ 
\hline
\multirow{2}{*}{0.4}  &\cellcolor{gray!25} regression tree  &\cellcolor{gray!25} \textbf{0.8}  &\cellcolor{gray!25} \textbf{5.4}  &\cellcolor{gray!25} 11.0  &\cellcolor{gray!25} 22.3  &\cellcolor{gray!25} \textbf{46.8}  &\cellcolor{gray!25} 141.0 \\ \cline{2-8}
 & knn  & 1.4  & 5.7  & \textbf{10.8}  & \textbf{20.3}  & 66.2  & \textbf{117.8} \\ 
\hline
\multirow{2}{*}{0.5}  &\cellcolor{gray!25} regression tree  &\cellcolor{gray!25} 0.9  &\cellcolor{gray!25} \textbf{4.7}  &\cellcolor{gray!25} \textbf{9.8}  &\cellcolor{gray!25} 21.4  &\cellcolor{gray!25} \textbf{51.5}  &\cellcolor{gray!25} 136.4 \\ \cline{2-8}
 & knn  & \textbf{0.6}  & 4.9  & 10.2  & \textbf{18.0}  & 68.6  & \textbf{117.5} \\ 
\hline
\multirow{2}{*}{0.6}  &\cellcolor{gray!25} regression tree  &\cellcolor{gray!25} \textbf{0.9}  &\cellcolor{gray!25} \textbf{4.5}  &\cellcolor{gray!25} 9.9  &\cellcolor{gray!25} 22.0  &\cellcolor{gray!25} \textbf{49.8}  &\cellcolor{gray!25} 131.6 \\ \cline{2-8}
 & knn  & 1.2  & 4.7  & \textbf{9.9}  & \textbf{17.3}  & 66.7  & \textbf{121.6} \\ 
\hline
\multirow{2}{*}{0.7}  &\cellcolor{gray!25} regression tree  &\cellcolor{gray!25} \textbf{0.8}  &\cellcolor{gray!25} 5.4  &\cellcolor{gray!25} 9.6  &\cellcolor{gray!25} 21.6  &\cellcolor{gray!25} \textbf{47.1}  &\cellcolor{gray!25} 133.6 \\ \cline{2-8}
 & knn  & 1.1  & \textbf{4.5}  & \textbf{9.2}  & \textbf{16.9}  & 65.2  & \textbf{108.9} \\ 
\hline
\multirow{2}{*}{0.8}  &\cellcolor{gray!25} regression tree  &\cellcolor{gray!25} 1.0  &\cellcolor{gray!25} 4.2  &\cellcolor{gray!25} 10.9  &\cellcolor{gray!25} 22.1  &\cellcolor{gray!25} \textbf{51.1}  &\cellcolor{gray!25} 131.8 \\ \cline{2-8}
 & knn  & \textbf{0.7}  & \textbf{3.8}  & \textbf{9.7}  & \textbf{16.7}  & 68.7  & \textbf{92.4} \\ 
\hline
\multirow{2}{*}{0.9}  &\cellcolor{gray!25} regression tree  &\cellcolor{gray!25} \textbf{0.8}  &\cellcolor{gray!25} 4.5  &\cellcolor{gray!25} 10.8  &\cellcolor{gray!25} 21.1  &\cellcolor{gray!25} \textbf{61.1}  &\cellcolor{gray!25} 131.6 \\ \cline{2-8}
 & knn  & 1.1  & \textbf{4.3}  & \textbf{8.5}  & \textbf{16.6}  & 66.1  & \textbf{90.4} \\ 
\hline
\multirow{2}{*}{1.0}  &\cellcolor{gray!25} regression tree  &\cellcolor{gray!25} 1.0  &\cellcolor{gray!25} 5.6  &\cellcolor{gray!25} 11.9  &\cellcolor{gray!25} 23.1  &\cellcolor{gray!25} 69.1  &\cellcolor{gray!25} 137.0 \\ \cline{2-8}
 & knn  & \textbf{0.6}  & \textbf{3.9}  & \textbf{8.0}  & \textbf{15.1}  & \textbf{48.3}  & \textbf{120.0} \\ 
\hline
\end{tabular}
\caption{
Absolute percentage error of the learned models on the testing set for different qualities.
The values in bold represent the best value among all regression methods.
The most interesting values are in the column for the $0.75$ quantile which represents the majority of the observations.
} \label{tab:exp/est/ape}
\end{table}

Figure \ref{fig:alg/estim/methods/objective_curves} shows an example of one observation with its measured processing time, approximated processing time function and estimated processing time functions (full and linear estimation).
The estimated processing time overestimates the measured processing time function which is better than if the processing time functions were underestimated.
To explain this, assume that there are two tasks $\task, \task'$ and the maximum normalized processing time of both tasks is $100\,ms$.
The estimated maximum normalized processing time of task $\task$ is $120\,ms$ (i.e. overestimation) and the estimated maximum normalized processing time of task $\task'$ is $80\,ms$ (i.e. underestimation).
Next, assume that for task $\task'$ the normalized processing time of $80\,ms$ relates to quality $0.8$.
When the scheduling system is not overloaded, i.e. the requested quality of the solution is 1, the actual quality of the solution for task $\task$ would be $1$ because the processing time is equal to $120\,ms$ which means that the algorithm is processing task $\task$ for more time than is required to get the solution of quality 1.
On the other hand, the actual quality of the solution  for task $\task$ would be $0.8$ because the processing time equals to $80\,ms$ which means that the algorithm is processing task $\task'$ for less time than is required to get the solution of quality 1.
Therefore, underestimation is undesirable if, most of the time, the scheduling system is not overloaded.

\subsection{Experiments with the scheduling system}
In the following experiments, the ability of the scheduling system to keep the average normalized lateness near 0 is analyzed.
The instances from the previous experiment are used.

In the experiments, the proposed bisection and individual control algorithms (see Section \ref{sec:alg/quality}) are compared to alternative control algorithms:
\begin{enumerate}
  \item \define{Max quality}: The quality control is disabled, i.e. the requested quality is set to 1 for all tasks.
    However, due to imprecision in the execution time estimation, the actual quality of the returned solutions does not have always to be 1.

  \item \define{Min quality}: The requested quality is set to some minimum quality which is denoted as $\minq$.

  \item \define{Random control}: Before a task is assigned to some resource, the requested quality of that task is fixed to some randomly sampled value from the uniform distribution $\unifdist(\minq, 1)$.
    
  \item \define{Naive}: When the overload is detected, the algorithm iteratively sets the lowest possible quality to tasks which are currently being processed and which has been processed for the longest time until the average normalized lateness is less or equal to 0.
\end{enumerate}

\subsubsection{Experimental setup}
In the experiments, $20$ heterogeneous resources are available to the scheduling system.
The speed of each resource was sampled randomly from the uniform distribution $\unifdist(1, 3)$.
The minimum quality $\minq$ was set to 0.2.

To generate the requested response time for each task, the instances of the tasks were split to disjoint bins with the width of 1 second by their maximum normalized processing time.
For each instance in each $bin_k$, the requested response time was randomly sampled from uniform distribution $\unifdist(1.5 k, 3k)$.
The advantage of this approach is that the requested response time is not directly dependent on the maximum processing time of the tasks.

To create a workload, the instances were split into three groups by their maximum processing time: 
\begin{enumerate}
  \item the maximum processing time is at most $2\,s$.

  \item the maximum processing time is greater than $2\,s$ and less than or equal to $20\,s$.

  \item the maximum processing time is greater than $20\,s$.
\end{enumerate}
For each group, six client applications (i.e. 18 client applications in total) were created which generated tasks at some rate and sent them to the scheduling system.
The inter-arrival time of the tasks, i.e. the difference between the arrival time of consecutive tasks, was sampled from the exponential distribution with a mean of $100\,ms$.
In total, 600 tasks were generated from the first group, 510 from the second group and 420 from the third group.
Therefore, the total number of generated tasks was $\ntasks=1430$.

\subsubsection{Evaluation of the quality control algorithms}
The results of each scenario, i.e. a combination of a control algorithm and an estimation method, are depicted in a figure consisting of three subfigures ordered from top to bottom: (i) the number of pending tasks in the system in each time unit, (ii) the normalized lateness of the solutions, and (iii) the quality of the solutions.

\begin{figure}[t]
  \centering
  \includegraphics[scale=0.6]{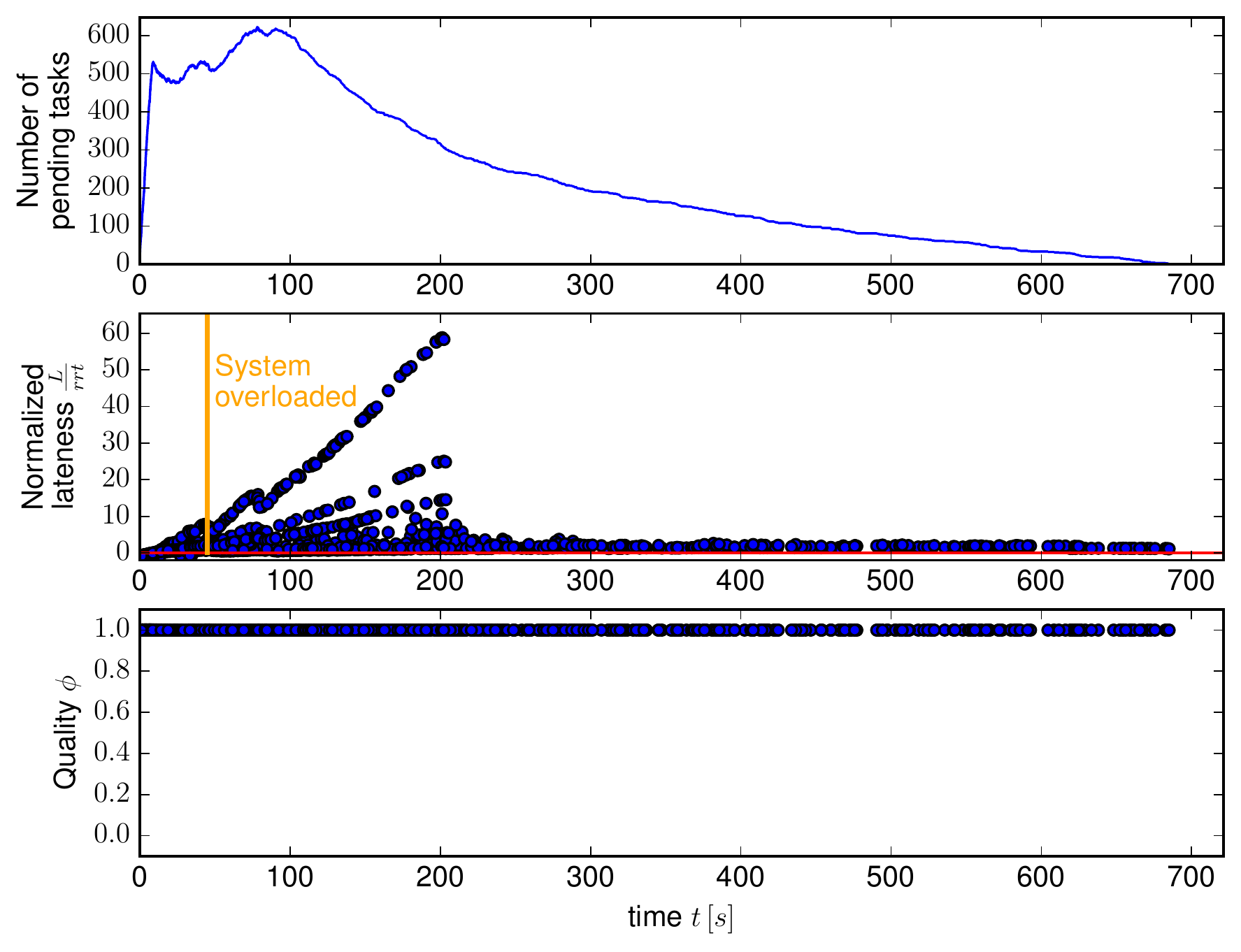}
  \caption{The dependence of the normalized lateness and the solution quality of tasks on the workload when the quality control is disabled (i.e. \textit{Max quality} control algorithm) and measured processing time is used.}
  \label{fig:exp/sim/ev_max}
\end{figure}

The results for the situation when the quality control is disabled (i.e. \define{Max quality} control algorithm) and the scheduler is using the measured processing time are shown in Figure \ref{fig:exp/sim/ev_max}.
It can be clearly seen that the scheduling system starts to be overloaded around $40\,s$ which results in a higher normalized lateness.
For some tasks, the lateness is 60 times larger than the requested lateness which is unacceptable.

\begin{figure}[H]
  \centering
  \includegraphics[scale=0.6]{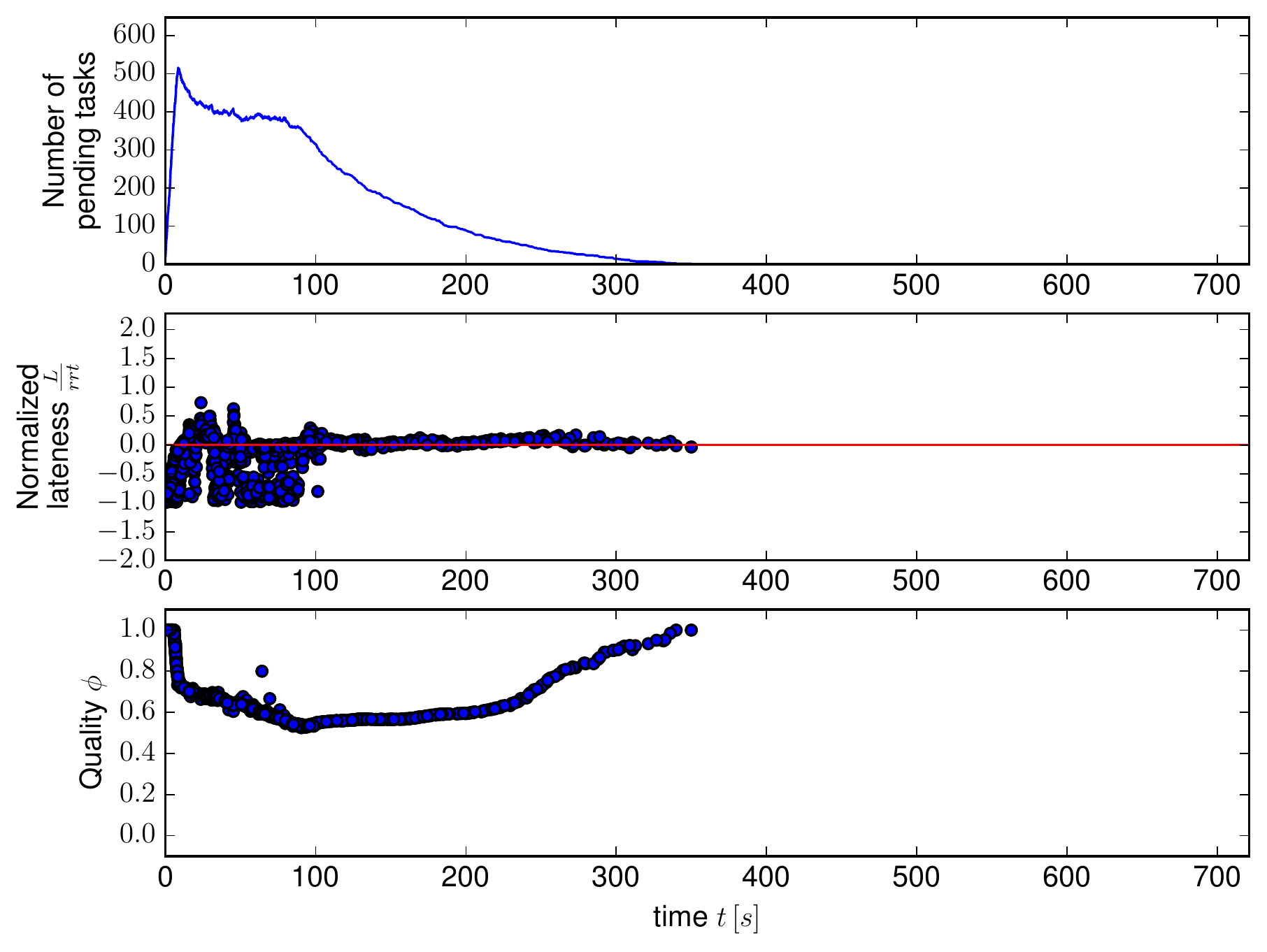}
  \caption{The dependence of the normalized lateness and the solution quality of the tasks on the workload when the bisection control is enabled and measured processing time is used.}
  \label{fig:exp/sim/ev_bin}
\end{figure}

\begin{figure}[H]
  \centering
  \includegraphics[scale=0.6]{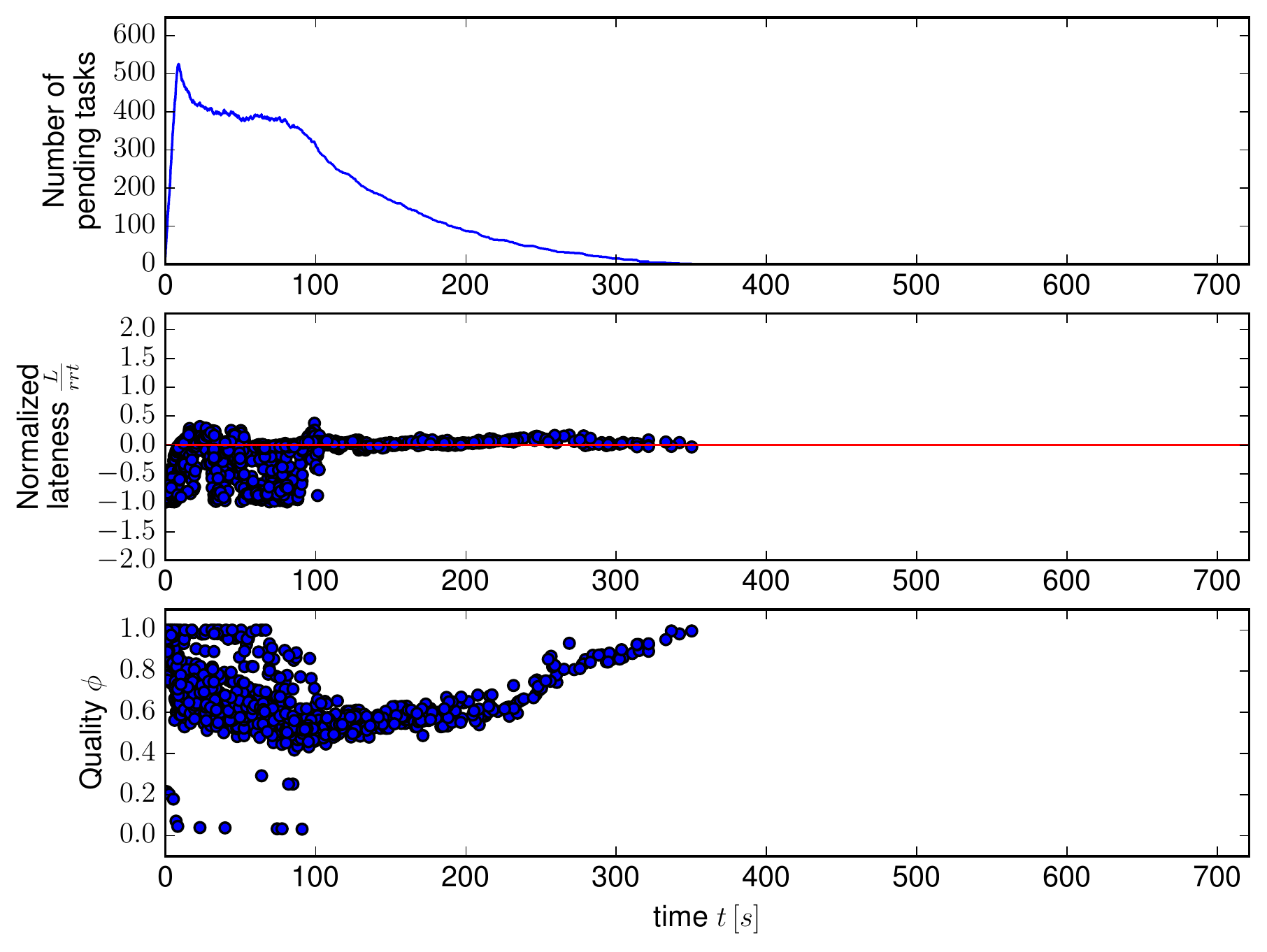}
  \caption{The dependence of the normalized lateness and the solution quality of the tasks on the workload when the bisection control is enabled and full estimation is used.}
  \label{fig:exp/sim/ev_bin_fp}
\end{figure}

The situation when the bisection control is enabled and the scheduler is using the measured processing time is demonstrated in Figure \ref{fig:exp/sim/ev_bin} (notice that the scale of the normalized lateness is different from Figure \ref{fig:exp/sim/ev_max}).
The control algorithm detects the overload and decreases the requested quality of the solutions so that the normalized lateness is around 0.
The requested quality of the solutions gradually increases as the overload decreases.
In Figure \ref{fig:exp/sim/ev_bin_fp}, the results for the same control method with the full estimation are shown.
Notice that the overall shape of the requested quality of the solutions is similar to Figure \ref{fig:exp/sim/ev_bin}.
Due to the inexact estimation, the spread of the quality is larger, however even in such a situation the response time of the tasks is still maintained at an acceptable level.

\begin{figure}[H]
  \centering
  \includegraphics[scale=0.6]{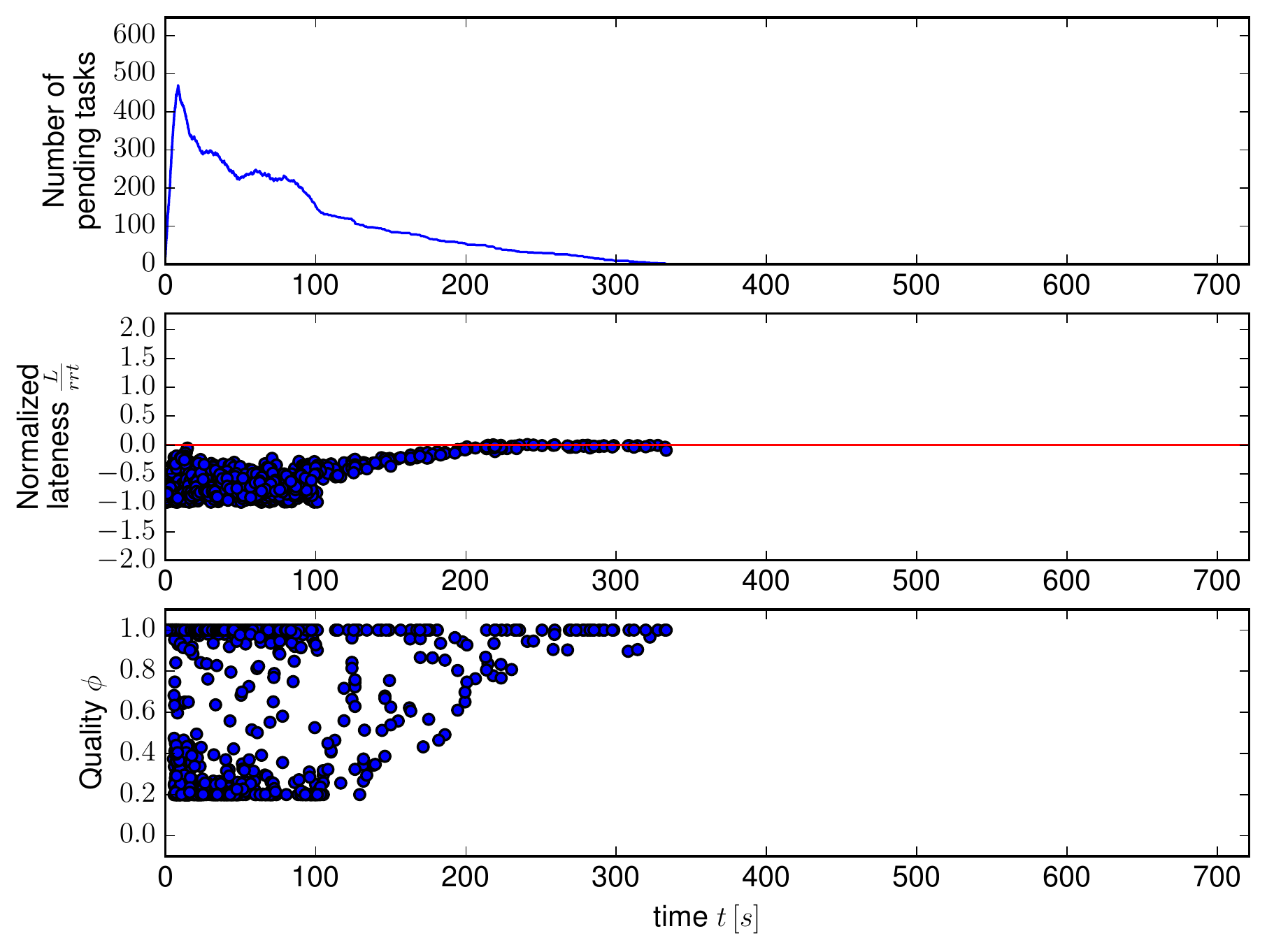}
  \caption{The dependence of the normalized lateness and the solution quality of the tasks on the workload when the individual control is enabled and measured processing time is used.}
  \label{fig:exp/sim/ev_ind}
\end{figure}

The results for the individual control with the measured processing time are shown in Figure \ref{fig:exp/sim/ev_ind} (the individual control with the full estimation is not included because the effect of the estimation is not clearly visible as with the bisection control).
The individual control algorithm is also able to keep the normalized lateness around 0.
From the quality of the solutions, it can be seen that the bisection and individual control algorithms behave differently.
Whereas the bisection control sets the same quality for all the tasks, the independent control can specifically address those tasks whose contribution to the average normalized lateness is the highest.
If the minimum quality were set to 0, the independent control would behave similarly to the admission control, i.e. it would drop the tasks with the highest contribution to the average normalized lateness.

\begin{figure}[H]
  \centering
  \includegraphics[scale=0.6]{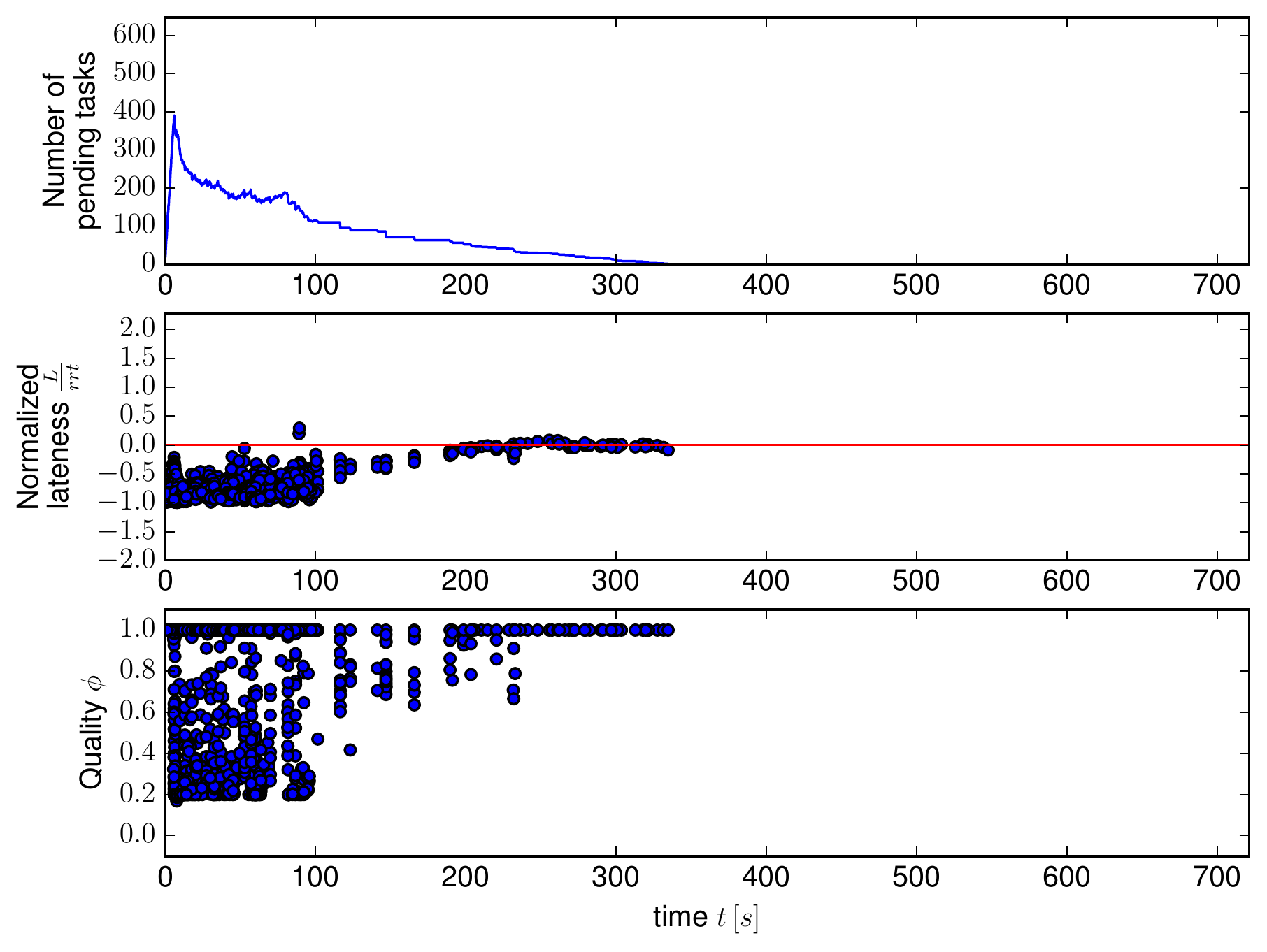}
  \caption{The dependence of the normalized lateness and the solution quality of the tasks on the workload when the naive control is enabled and measured processing time is used.}
  \label{fig:exp/sim/ev_naive}
\end{figure}

The last Figure \ref{fig:exp/sim/ev_naive} illustrates the situation when the naive control with the measured processing time is used.
Similarly to the individual control, the naive control sets the quality for each task and is also able to keep the response time around the acceptable level.

\subsubsection{Overall comparison}
\begin{figure}[H]
  \centering
  \includegraphics[scale=0.7]{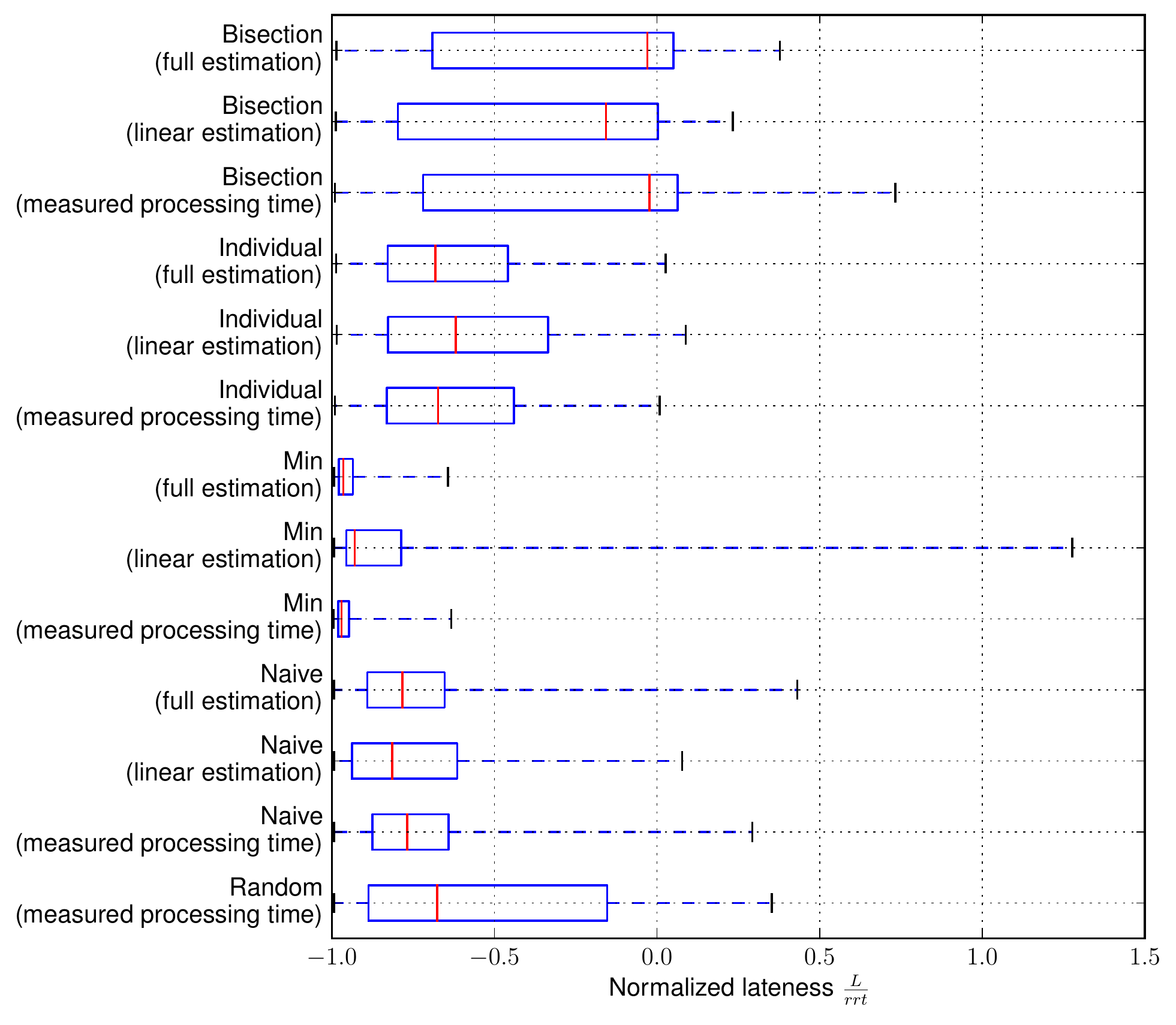}
  \caption{
    Comparison of various quality control and estimation methods by normalized lateness.
    The edges of the boxes represent the lower and upper quartile, the red vertical line inside each box is median and the whiskers extend from the minimum to maximum value.
  }
  \label{fig:exp/sim/nl}
\end{figure}

\begin{figure}[H]
  \centering
  \includegraphics[scale=0.7]{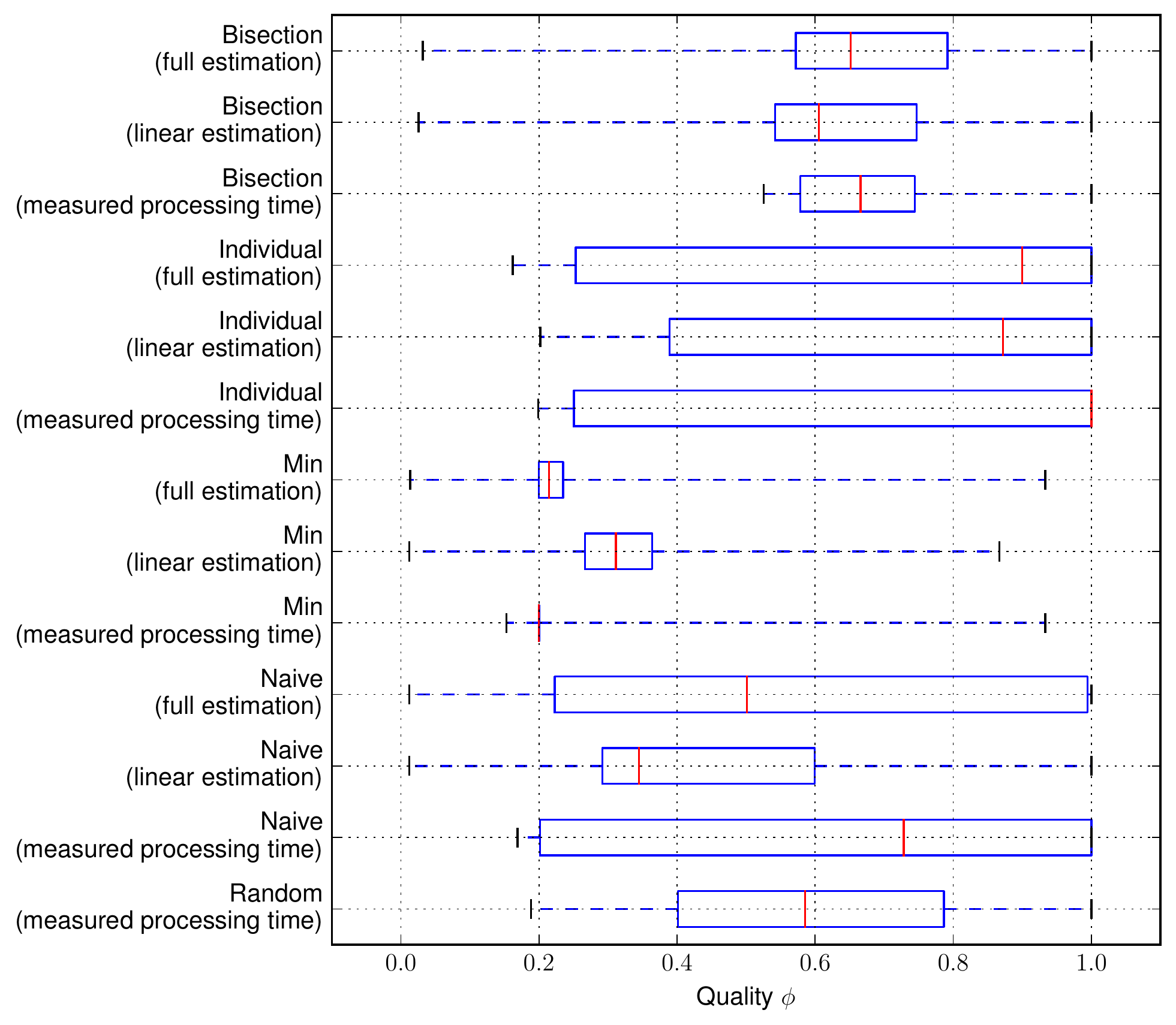}
  \caption{
    Comparison of various quality control and estimation methods by solution quality.
    The edges of the boxes represent the lower and upper quartile, the red vertical line inside each box is median and the whiskers extend from the minimum to maximum value.
  }
  \label{fig:exp/sim/q}
\end{figure}

Figures \ref{fig:exp/sim/nl} and \ref{fig:exp/sim/q} show the spread of the normalized lateness and the quality of the solutions, respectively.
Observe that the maximum normalized lateness for the bisection control, when the full estimation is used, is less than 0.38 and in 75\% of the cases it is less than 0.051.
Although these values are slightly higher than 0, it is much better than the maximum normalized lateness of the \define{Max quality}.
However, the accuracy of the estimation affects the spread of the quality.
The minimum solution quality for the bisection control when the full estimation is used is considerably lower than when the scheduling system is using the measured processing time.
However, in most cases (i.e. 75\%), the solution quality is sufficiently high.

For the individual control with the full estimation, the vast majority of the tasks have a lateness of less than 0.03 which is a great result considering that this result was not obtained at the cost of significantly decreasing the quality of the solutions.
The quality of the solutions is also affected by the estimation but, as with the bisection control, the difference is small in comparison with the situation when the measured processing time is used.
In comparison with the naive control, the individual control has much higher median of the solution quality.
In the case of the full estimation, the difference in the median of the solution quality is almost 0.3.
We note that even though the normalized lateness of the naive control is smaller compared to the individual control, the difference is not relevant because the normalized lateness of both control methods is negative - we aim for non-positive average normalized lateness.

The reason why the quality for the \define{Min quality} with the measured processing time has a large nonzero spread is because the smallest time unit is set to $1\,ms$ which influences the instances with a small processing time.
For example, consider a task instance with a linear processing time function and with the maximum normalized processing time of $100\,ms$.
If the requested quality is set to 0.2, the normalized processing time equals to $20\,ms$.
If the scheduling system decides to assign that task to a resource $\resource$ with a speed of $\speed* = 100$, it computes that the estimated processing time of that task is
\begin{equation}
  \ceil*{\frac{\nptfn(\instance, 0.2)}{\speed*}} = \ceil*{\frac{20}{100}} = \ceil{0.2} = 1\,,
\end{equation}
i.e. resource $\resource$ should process the task for $1\,ms$.
After converting that processing time to the normalized processing time, i.e. $100\,ms$, it is obvious that the quality of the solution is 1 even though the requested quality is set to 0.2.

\begin{figure}[H]
  \centering
  \includegraphics[scale=0.6]{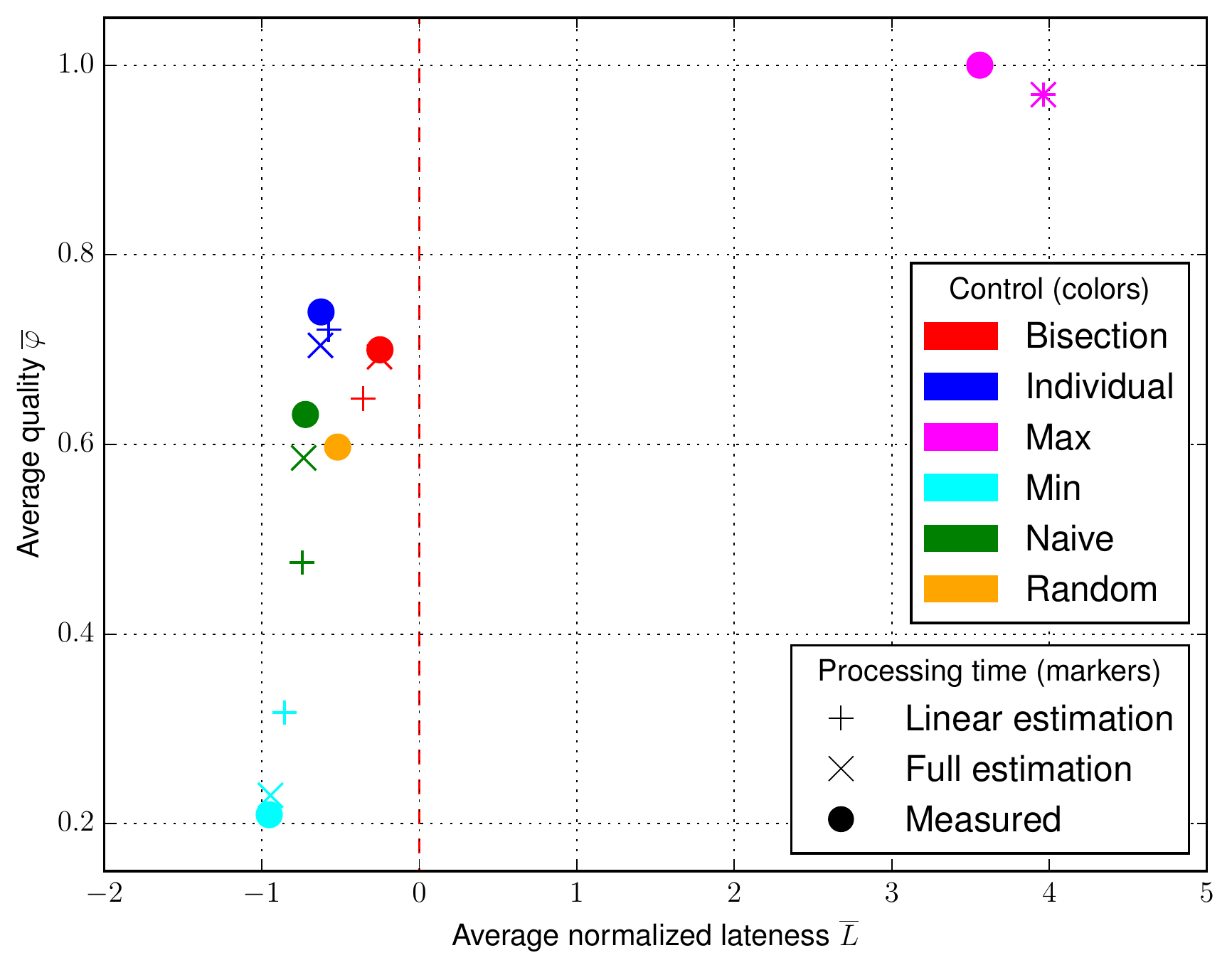}
  \caption{
    Comparison of various quality control and estimation methods by the average quality of the solutions and the average normalized lateness.
    The markers for random control are overlapping.
  }
  \label{fig:exp/sim/anl_vs_aq}
\end{figure}

The overall comparison of various quality control algorithms and estimation methods is shown in Figure \ref{fig:exp/sim/anl_vs_aq}.
The methods are compared by plotting the average quality of the solutions and the average normalized lateness.
It can be seen that the full estimation does not significantly influence the average values, i.e. the scheduling system is robust against small errors in the estimation.
When the bisection control is used, the average quality of the solutions for the full estimation achieves the highest quality, whereas the linear estimation is the worst.
The full estimation is only slightly worse than using the measured processing time.
When the individual control is used, the difference between the full and linear estimation is small because the individual control, according to Equation \ref{eq:nalat_expand}, uses only the estimation of the maximum normalized processing time.
The difference arises from the fact that the scheduling policy uses the linear and full estimation in their original form.
The result of the random control is not surprising because the mean of the uniform distribution $\unifdist(\minq, 1) = \unifdist(0.2, 1)$ is $\mu = 0.6$.

Figure \ref{fig:exp/sim/anl_vs_aq} also shows that the total solution quality is approximately the same for the individual and bisection control.
The difference of these methods is how the quality is distributed among the tasks.
Whereas the bisection control distributes the quality uniformly among the pending tasks, the individual control can compress the processing time of those tasks that have a large contribution to the average normalized lateness.
The effect of this is that more solutions have a quality of 1 at the cost of having a few solutions of minimum quality.
If the minimum quality were set to 0, the individual control would compress some tasks completely and, therefore, the individual control would behave similarly as an admission control.
Such behavior may not be desirable in some applications and, therefore we cannot declare that the individual control is better than the bisection control even though the numeric results are in favor of the individual control.

To compare the naive and the individual control, we note that the difference in the average solution quality between the corresponding estimation methods is at least $10\%$ in favor of the individual control.
Since the objective functions on the application layer (i.e. the objective functions of the task instances) relate to the quality of the solutions, such difference can be significant, e.g. consider an objective function denoting the saved money achieved by a solution.
Moreover, in other scenarios the naive control may perform much worse, e.g. consider a situation when short task $1$ and two much longer $\{2,3\}$ arrive in the system at the same time.
Assume that the scheduler creates schedule $(1,2,3)$ on one resource and the last task $3$ is missing its requested due date.
The naive control decreases the quality of the short task $1$ even though it has a negligible effect on the completion time of the longer tasks and, therefore, the requested due date of task $3$ is still being missed.
After the solution of task $1$ is received, the quality is recomputed.
The naive control decreases the quality of task $2$, which considerably affects the completion time of task $3$ and the requested due date of task $3$ is satisfied.
On the other hand, the individual control detects that the congestion is caused by task $2$ leaving the quality of task $1$ intact.

The question may arise about the low normalized lateness of the individual control: could be the quality of solution increased so that the  normalized lateness is closer to zero? Recall that the average normalized lateness in the individual control is computed per resource.
Consider an example with  two resources $\{1,2\}$ of speed 1 and three tasks $\{1,2,3\}$ with parameters from Table \ref{tab:exp/sim/less_zero}.
If the average normalized lateness is computed over all resources for these tasks, then it is equal to $-0.1296$.
However, if the average normalized lateness is computed for each resource independently, then it  is equal to $0.056$ for resource 1 and for resource 2 it is equal to $-0.5$.
Since all tasks are running to the maximum quality and the overall average normalized lateness is less than 0, the bisection control would do nothing whereas the individual control would compress the tasks on resource $2$ so that the average normalized lateness is closer to zero.
By compressing the tasks, the solutions are received earlier and, therefore, the average normalized lateness of all the tasks in the experiments for the individual control is less than for the bisection control.
The average quality of both control algorithm is, in the end, similar because the individual control is able to compress a single bottleneck task without affecting the rest of the tasks.

\begin{table}[H]
  \centering
  \begin{tabular}{|ccccccc|}
  \hline
  $\task$ & $\nptfn(\instance*, 1)$ & $\reqresp*$ & $\arrival*$ & $\rallocsol*$ & $\allocsol*$ & $\compfn(\task, \rallocsol*, \allocsol*, 1)$
  \\ \hline
  1 & 10 & 10 & 0& 1&0 &  10 \\ \hline
  2 & 90 & 90 & 0& 1&10 & 100\\ \hline
  3 & 10 & 20 & 0& 2&0 &  10 \\ \hline
  \end{tabular}
  \caption{Parameters of the tasks for example explaining why the average normalized lateness of the individual control is less than 0.}
  \label{tab:exp/sim/less_zero}
\end{table}

\subsubsection{Performance of the algorithms}
\begin{figure}[H]
  \centering
  \includegraphics[scale=0.57]{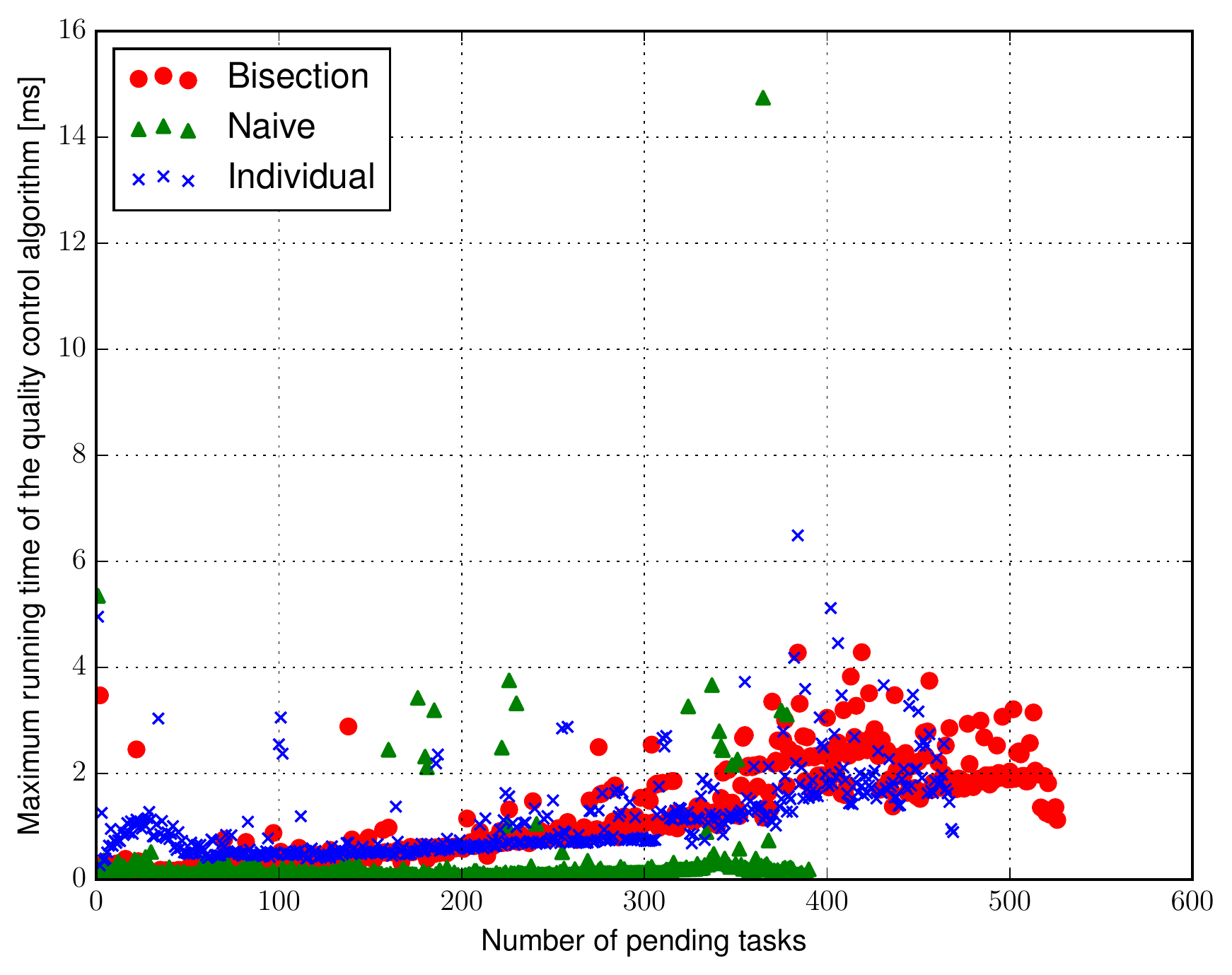}
  \caption{
    Comparison of the control algorithms by maximum running time.
    For each number of pending tasks there is, at most, one point (this is not obvious due to the density of the points).
  }
  \label{fig:exp/sim/pending_vs_qtime}
\end{figure}

\begin{figure}[H]
  \centering
  \includegraphics[scale=0.57]{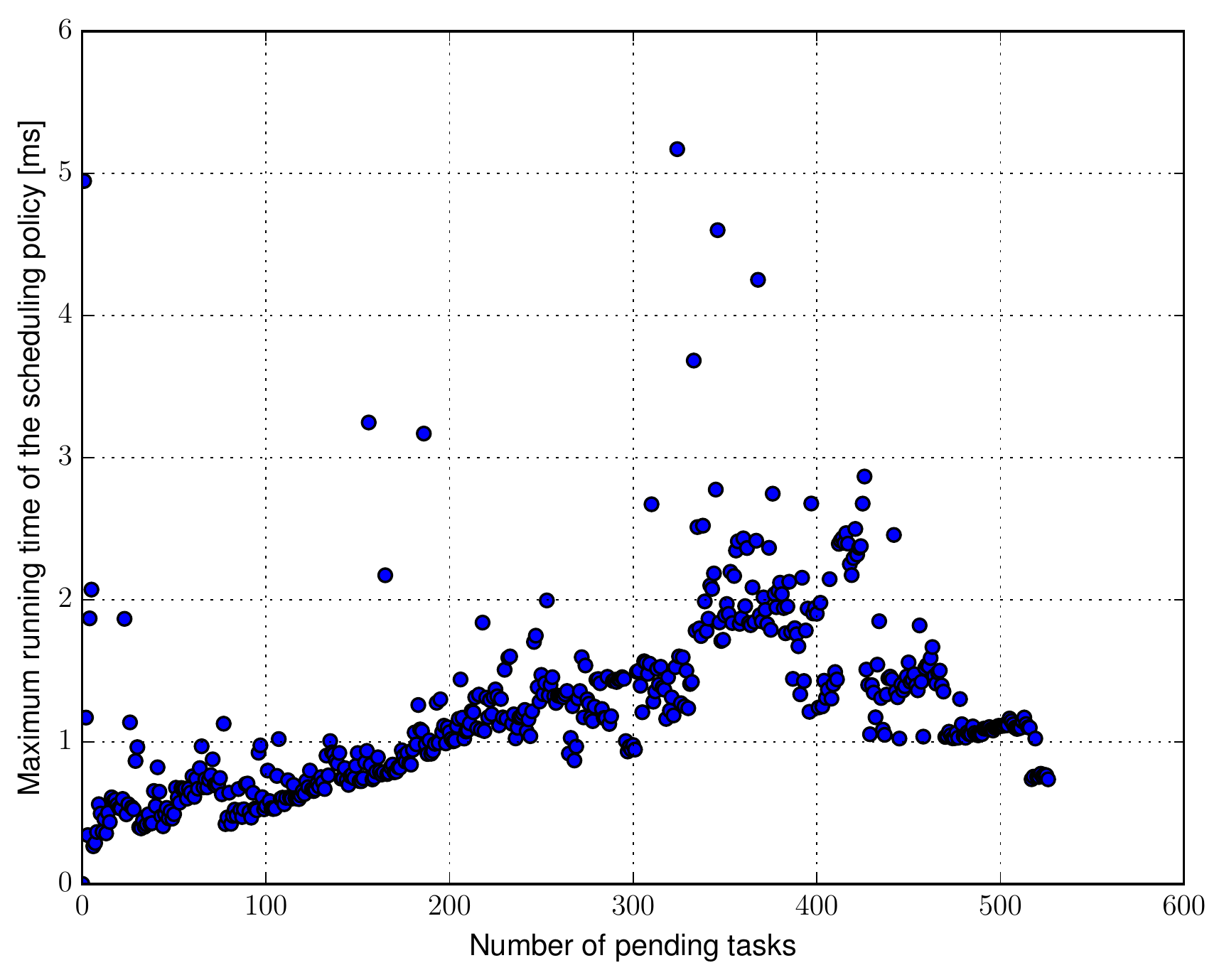}
  \caption{
    Maximum running time of the scheduling policy (since all control algorithms are using the same scheduling policy, the points are not distinguished by the control algorithms).
    For each number of pending tasks there is, at most, one point (this is not obvious due to the density of the points).
  }
  \label{fig:exp/sim/pending_vs_schtime}
\end{figure}

The measured dependence between the number of pending tasks and the running time of the proposed control algorithms is presented in Figure \ref{fig:exp/sim/pending_vs_qtime}.
The values for each number of pending tasks are obtained by choosing a maximum running time among all estimation methods.
The results of all methods show that the running time of the quality computation is quite low even for a high number of pending tasks and, therefore, negligible for the whole system.
Similarly, Figure \ref{fig:exp/sim/pending_vs_schtime} shows the dependency between the number of pending tasks and the running time of the scheduling policy. Again, the running time is negligible.

The reason for such small running times of the individual control algorithm is that the case when all pending tasks are assigned to one resource is not common in typical problem instances.
The number of pending tasks on each resource are rather balanced which leads to a lower running time.

%% file: conclusion.tex
\section{Conclusion}
\label{sec:conc}

In this work, a problem of scheduling tasks in a heterogeneous environment is considered, i.e. the scheduling system has to determine the assignments of the tasks to the available computational resources.
The goal of the scheduling system is to handle overload situations in which many computationally intensive tasks arrive in the system.
In such a situation, the response time of the server increases considerably and this leads to user dissatisfaction.

The first characteristic of the tasks is that they are solved by anytime algorithms, i.e. the quality of the solutions depends on the processing time of the tasks.
The second characteristic is that the relationship between the solution quality and the processing time of the tasks is not known \textit{a priori}.
Therefore, to make intelligent decisions the scheduling system has to estimate it.

Since the problem in its entirety is not already addressed in the existing literature, we proposed a modular system architecture, two efficient quality control algorithms and a procedure for estimating the processing time functions of the tasks.
Both quality control algorithms exploit the anytime property, i.e.  when the overload is detected, the algorithms decrease the requested quality of the solutions, thus, decreasing the response time of the server.
The algorithms differ in how the requested quality is controlled; the \define{bisection control} sets one global quality for all tasks, whereas the \define{individual control} controls the quality of each task independently.
The algorithmic complexities of the bisection and individual control algorithms are $\mathcal{O}\left(maxIters \cdot \left(\nresources + \npendtasks\right)\right)$ and $\mathcal{O}\left(\npendtasks \log{\npendtasks}\right)$, respectively ($\nresources$ is a number of resources and $\npendtasks$ is a number of tasks in the system).
Thanks to the low algorithmic complexity, the system can be used in online environments and is able to handle even hundreds of tasks.

To estimate the relationship between the solution quality and the processing time we introduced a procedure based on regression analysis.
The measured relationships from the previously completed tasks are approximated by a piecewise linear function with ten segments whose endpoints are used to train the regression methods.
The regression methods then estimate the processing time functions of the incoming tasks.

The design of the system is evaluated on a real client-server application from a domain of \textit{personnel rostering}.
The experiments show a huge decrease in the response time when either of the proposed quality control algorithms is used.
The proposed algorithms also outperformed a simple control approach which always stop a task that was running for the longest time.
The experiments also show that both control algorithms are robust against up to 20\% error in the estimation, i.e. the results in the average solution quality and the average normalized lateness are not significantly different if the estimation is used.
The positive results verify the validity of our approach. 

Although the scheduling and quality control algorithms can be used in any domain where the anytime tasks are scheduled, future research could focus on generalizing the estimation methods to domains other than personnel rostering.
Currently, we require that the relevant features of the task instances are identified manually so that the estimation can be performed.
However, the scheduling system could perform automatic \define{feature learning} based on the description of the task instances.
Moreover, the estimation error could be further decreased by using different regression methods such as Deep Learning.

Future research could also focus on extending the individual control algorithm with non-linear processing time functions.
We propose two ideas how to extend this algorithm.
The first method splits each task along the linear segments of the piecewise linear function (see Section \ref{sec:estim}).
The original task is replaced by the newly created tasks.
If each task consists of ten segments, then the complexity is $\mathcal{O}\left(10\npendtasks \log\left(10\npendtasks\right)\right)$, however, to ensure the correct behavior of the algorithm, the slopes of each segment of one task has to be non-decreasing.
The second method does not split the tasks but uses only the last segments of the piecewise linear functions to compute the weights.
The task with the highest weight is compressed only up to the starting endpoint of the last segment, then the last segment of that task is thrown away and the weight of that task is recomputed using the one before the last segment.
The advantage of this method is that there is no assumption of the shape of the piecewise linear functions as with the previous method.
However the complexity of this method is $\mathcal{O}\left(\npendtasks \cdot \npendtasks\right)$.

%% file: acknowledgment.tex
\section*{Acknowledgment}
\label{sec:ack}
This work was supported by the ARTEMIS initiative funded by the European Commission under the project DEMANES 295372.

%% file: appendix.tex
\afterpage{\clearpage}
\setcounter{thm}{0}
\section{Total complexity of computing Algorithm \ref{alg:alg/individual}
for all resources}
\label{sec:app/comp}

\begin{lem}
  \label{lem:app/comp/task_one_resource_worse}
  Let $x_1,x_2, \dots, x_n \in \pnaturalset, n \ge 2$. Then it holds that
  \begin{equation}
    \left( \sum_{i = 1}^n x_i \right) \log{\left(\sum_{i = 1}^n x_i\right)} > \sum_{i =
    1}^n x_i \log{x_i}\,.
  \end{equation}
\end{lem}
\begin{pf}
  We start by noting that
  \begin{equation}
    \frac{\sum_{j = 1}^n x_j}{x_i} > 1\,,\quad i = 1, \dots, n
  \end{equation}
  We use this result to prove Lemma \ref{lem:app/comp/task_one_resource_worse}
  \begin{equation}
    \begin{aligned}
          \left( \sum_{i = 1}^n x_i \right) \log{\left(\sum_{i = 1}^n x_i\right)}
          &>
          \sum_{i = 1}^n x_i \log{x_i}
      \\
          \log{\left( \sum_{i = 1}^n x_i \right)}^{\left( \sum_{i = 1}^n x_i \right)}
          &>
          \log{\left( \prod_{i = 1}^n x_i^{x_i} \right)}
      \\
          \left( \sum_{i = 1}^n x_i \right)^{\left( \sum_{i = 1}^n x_i \right)}
          &>
          \prod_{i = 1}^n x_i^{x_i}
      \\
          \prod_{i = 1}^n \left( \sum_{j = 1}^n x_j \right)^{x_i}
          &>
          \prod_{i = 1}^n x_i^{x_i}
      \\
          \prod_{i = 1}^n \left( \frac{\sum_{j = 1}^n x_j}{x_i} \right)^{x_i}
          &>
          1
    \end{aligned}
  \end{equation}

  \qed
\end{pf}

\begin{thm}
  \label{thm:app/comp/total_comp}
  The total worst case complexity of computing Algorithm \ref{alg:alg/individual} for all resources is $\mathcal{O}\left(\npendtasks \log{\npendtasks}\right)$.
\end{thm}
\begin{pf}
  In Section \ref{sec:alg/quality/ind}, the complexity of computing Algorithm \ref{alg:alg/individual} for one resource has been discussed already.
  To prove Theorem \ref{thm:app/comp/total_comp}, it has to be shown that the worst case is when all pending tasks are assigned to only one resource.

  Assume that not all pending tasks are assigned to one resource, i.e.
  \begin{equation}
    \exists \, \resource', \resource'' \in \resourceset: \resource' \not= \resource''
    \,\wedge \,
    \ntasks[\resource'] \ge 1
    \,\wedge \,
    \ntasks[\resource''] \ge 1 \,.
  \end{equation}
  This also implies that there are at least two pending tasks, i.e. $\npendtasks \ge 2$.
  The total complexity of computing Algorithm \ref{alg:alg/individual} for all resources is
  \begin{equation}
    \sum_{\resource \in \resourceset: \ntasks* \ge 1} \ntasks*
    \log{\ntasks*} \,.
  \end{equation}
  From Lemma \ref{lem:app/comp/task_one_resource_worse} it holds that
  \begin{equation}
    \label{eq:app/comp/after_lemma}
    \left(\sum_{\resource \in \resourceset: \ntasks* \ge 1} \ntasks* \right)
    \log{\left(\sum_{\resource \in \resourceset: \ntasks* \ge 1} \ntasks*\right)}
    >
    \sum_{\resource \in \resourceset: \ntasks* \ge 1} \ntasks* \log{\ntasks*}
  \end{equation}
  Because all tasks have to be assigned to some resource, it holds that
  \begin{equation}
    \label{eq:app/comp/npendtasks_equal_sum}
    \npendtasks = \sum_{\resource \in \resourceset: \ntasks* \ge 1}
    \ntasks*\,.
  \end{equation}
  By substituting $\npendtasks$ into inequality \eqref{eq:app/comp/after_lemma} we get
  \begin{equation}
    \npendtasks \log{\npendtasks} > \sum_{\resource \in \resourceset:
    \ntasks* \ge 1} \ntasks* \log{\ntasks*}\,.
  \end{equation}
  This inequality proves that the worst case is when all pending tasks are assigned to only one resource.

  \qed
\end{pf}